# Evidence for supramolecular dynamics of non-hydrogen bonding polar van der Waals liquids

Shalin Patil,[1] Catalin Gainaru,[2] Roland Böhmer,[3] and Shiwang Cheng[1,*]

[1] *Department of Chemical Engineering and Materials Science, Michigan State University, East Lansing, MI 48824, United States*

[2] *Chemical Sciences Division, Oak Ridge National Laboratory, Oak Ridge, Tennessee, 37831, United States*

[3] *Fakultät für Physik, Technische Universität Dortmund, Dortmund 44227, Germany*

**Abstract**

Non-hydrogen bonding van der Waals liquids with dipole-dipole interactions are typically viewed as non-associative and not considered able to sustain large supramolecular structures. Combining broadband dielectric spectroscopy (BDS) and rheology, we demonstrate the supramolecular formation in a group of non-hydrogen bonding van der Waals liquids, i.e. 1-bromo-2-ethylhexane, 1-chloro-2-ethylhexane, and 1-bromo-3,7-dimethyloctane. BDS shows an emergence of a Debye-like process slower than their structural relaxation, which follows super-Arrhenius temperature dependence. Meanwhile, rheological measurements reveal a noticeable dynamical separation between the terminal relaxation and the structural rearrangements. Interestingly, the rheological terminal time agrees remarkably well with the dielectric Debye-like relaxation time, pointing to a strong coupling between the terminal flow and the supramolecular dynamics of these van der Waals liquids. These results highlight the role of intermolecular dipole-dipole interaction on the structure and slow dynamics of van der Waals liquids.

---

[*] Corresponding Author. Email Address: chengsh9@msu.edu

## 1. Introduction

Intermolecular dipole-dipole interactions are the key for the rich macroscopic properties of molecular liquids, including their dielectric constant, refractive index, viscosity, glass transition, melting point, and boiling point.[1-2] However, understanding how these secondary intermolecular interactions affect materials properties remains a topic of active research.[3-6] In particular, van der Waals liquids have been viewed as non-associative and have been treated as model systems to test liquid state theories, phase transitions, and to understand the dynamics of supercooled liquids or the glass transition.[4, 7] Recent studies of broadband dielectric spectroscopy (BDS), dynamic light scattering (DLS), and nuclear magnetic resonance (NMR) spectroscopy suggest a Debye-like relaxation of molecular liquids slower than their structural relaxation for glass transition, including the hydrogen bonding liquids and some polar van der Waals liquids.[8-12] The new observations attract active discussions on the physical origin of the Debye-like relaxation processes of molecular liquids, including the dipole-dipole cross-correlation at the molecular level.[9, 13-16] On the other hand, Debye-like processes of some hydrogen bonding liquids, such as the monohydroxy alcohols and their mixtures, have been actively studied in the past and have been attributed to a type of supramolecular dynamics.[17-23] These seemingly different opinions motivate new discussions on the origin and the characteristics of the Debye-like relaxation of molecular liquids, especially for the non-hydrogen bonding van der Waals liquids.

In this work, we investigate the dynamics of a group of non-hydrogen bonding van der Waals liquids with dipole-dipole interactions, i.e., 1-bromo-2-ethylhexane (2E1Br), 1-chloro-2-ethylhexane (2E1Cl), and 1-bromo-3,7-dimethyloctane (3,7D1OBr). BDS and rheology have been applied to study the dynamics and viscoelastic properties of these van der Waals liquids. BDS measurements of all three liquids demonstrate more than one active relaxation processes that exhibit super-Arrhenius temperature dependences with the one at low frequencies exhibiting Debye-like dielectric dispersion. Linear rheology, at the same time, shows a large separation in time scales between structural relaxation and the terminal flow. Interestingly, both the structural relaxation time and the terminal flow time from rheology display correspondences to dielectric structural relaxation and the Debye-like process respectively, indicating the presence of collective supramolecular modes and that these supramolecular dynamics are dielectrically active. These results suggest polar van der Waals liquids can form relatively large supramolecular structures that lead to the separation between the time scales for flow and the structural relaxation, and give the Debye-like relaxation processes.

## 2. Materials and Methods

### 2.1. Materials

1-bromo-2-ethylhexane (2E1Br, Sigma-Aldrich, Product No. 249416) and 1-bromo-3,7-dimethyl octane (3,7D1OBr, Sigma-Aldrich, Product No.533904) were obtained from Sigma-Aldrich. The molecular structures of these liquids are represented in **Figure 1**, where a C-X (X = Br or Cl) bond carries a permanent dipole moment $\mu \approx 1.56$ D (for C-Cl) and $\mu \approx 1.48$ D (for C-Br).[24] 2E1Br and 2E1Cl have eight carbon atoms and 3,7D1OBr has ten carbons in the alkyl group. The mass density of the 2E1Br was measured through quantifying its volume of a given mass confined between two parallel plates at different temperatures in a rheometer (for details see **Supplementary Materials** (**SM**) and **Figure S1**). To remove the possible influence of impurities on the measurements, 2E1Br and 3,7D1OBr were purified through distillation as confirmed through NMR spectroscopy measurements (**Figure S2** and **S3** of the **SM**). It is worth noting that 2E1Br without purification gives a pronounced additional relaxation process at low frequencies, while the distilled 2E1Br does not (**Figure S4** of the **SM**). At the same time, the BDS spectra of 3,7D1OBr remain almost the same before and after purification. Hence, the as purchased molecular liquids might include impurities that can affect the BDS spectra. To eliminate the influence of impurities, all the results of 2E1Br and 3,7D1OBr are from samples after sample purification, and the results of 2E1Cl are replots of Ref. 20.

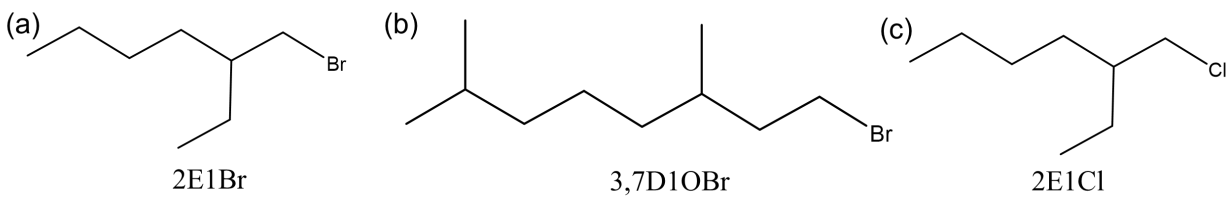

**Figure 1**. The molecular structure of **(a)** 1-bromo-2-ethylhexane (2E1Br), **(b)** 1-bromo-3,7-dimethyl octane (3,7D1OBr), **(c)** 1-chloro-2-ethylhexane (2E1Cl).

## 2.2. Methods

### 2.2.1. Broadband Dielectric Spectroscopy

Broadband dielectric spectroscopy (BDS) was employed to quantify the dynamics of these molecular liquids. Specifically, 2E1Br or 3,7D1OBr were sandwiched by two gold electrodes with a diameter of 20 mm. A hollow Teflon spacer of a thickness 0.14 mm and an inner diameter of 16 mm and an outer diameter of 20 mm was placed between the two gold electrodes to prevent shortage. The sandwiched samples were then loaded into the ZGS sample holder of a Novocontrol Concept-40 system with an Alpha-A impedance analyzer and a Quatro Cryosystem temperature controller. The temperature system has an accuracy of ±0.1 K. In all measurements, we scan the frequencies from $10^7$ Hz to $10^{-2}$ Hz and applied a root-mean-squared AC voltage of 1.0 V. Dielectric measurements were performed at both cooling and heating from 293 K to 183 K at an interval of 10 K upon cooling, and from 183 K to 133 K at an interval of 5 K. A thermal annealing of 1,200 s was applied before each measurement to assure thermal equilibrium. We compare the spectra upon heating and cooling to confirm the reproducibility of the results.

### 2.2.2. Rheology

Linear viscoelastic measurements were conducted through small amplitude oscillatory shear (SAOS) on an Anton Paar (MCR302) rheometer. The temperature was regulated by a CTD600 oven equipped with a liquid nitrogen evaporation unit (EVU20), which has an accuracy of ±0.1 K. A pair of parallel plates of 4 mm in diameter was utilized in the measurements and the strain amplitude was varied from ~0.01% at temperatures close to glass transition temperature, $T_g$, to ~5% at temperatures when the terminal regime was accessed. The sample thickness was maintained at 1 mm during the measurements and the angular frequency was set as 628 - 0.1 rad/s. The samples were loaded at room temperature and cooled immediately to close to $T_g$ for the measurements. The measurements covered from $T_g$ to temperatures where the terminal regions were fully captured.

### 2.2.3. The Kirkwood-Fröhlich factor, $g_k$

The Kirkwood-Fröhlich equation was used to estimate the Kirkwood-Fröhlich factor, $g_k$, which provides a characterization of the orientational correlation of the dipoles. In particular, $g_k > 1$ indicates the presence of parallel orientations of dipoles, and $g_k < 1$ indicates the presence of anti-parallel alignment of dipoles. The Kirkwood-Fröhlich equation is given by[25-27]

$$g_k = \frac{9(\varepsilon_s - \varepsilon_\infty)(2\varepsilon_s + \varepsilon_\infty)k_B M \varepsilon_0 T}{\varepsilon_s(\varepsilon_\infty + 2)^2 \mu^2 \rho N_A} \quad (1)$$

where, $\varepsilon_s$ is the static dielectric constant, $\varepsilon_\infty$ is the high frequency dielectric constant, $k_B$ is the Boltzmann constant, $M$ is the molar mass, $\varepsilon_0$ is the permittivity of vacuum, $\mu$ is the dipole moment, $\rho$ is the mass density, $N_A$ is Avogadro's number.

## 3. Results and Discussions

### 3.1 The Broadband Dielectric Spectroscopy (BDS)

**Figure 2** presents the representative spectra of the dielectric storage permittivity, $\varepsilon'(\omega)$ (**Figure 2a**), the derivative spectra, $\varepsilon'_{der}(\omega) = -\frac{\pi}{2}\frac{\partial \varepsilon'(\omega)}{\partial \ln \omega}$ (**Figure 2b**), and the loss permittivity, $\varepsilon''(\omega)$ (**Figure 2c**), of 2E1Br at temperatures $T = 133K, 138K, 143K, 148K, 153K$, and $158K$, where $\omega$ is the angular frequency. A dominant dielectric relaxation can be identified at each temperature from the step of $\varepsilon'(\omega)$ and from the characteristic peaks of both $\varepsilon'_{der}(\omega)$ and $\varepsilon''(\omega)$. Normalizing $\varepsilon''(\omega)$ by the values of its characteristic peak frequency $\omega_p$, $\varepsilon''(\omega)/\varepsilon''(\omega_p)$ vs $\omega/\omega_p$, gives a nice overlapping in the dielectric response at the low-frequency region and an increment in spectral broadening at high frequencies upon cooling (**Figure 2d**). Similar broadening of the high-frequency side of the dielectric spectra can also be observed in the derivative spectra (**Figure S5** in the **SM**). The high-frequency broadening of the dielectric spectra suggests the presence of an additional relaxation process faster than the main peak. To facilitate the discussion, we designate the relaxation processes from low to high frequencies as Processes I and II.

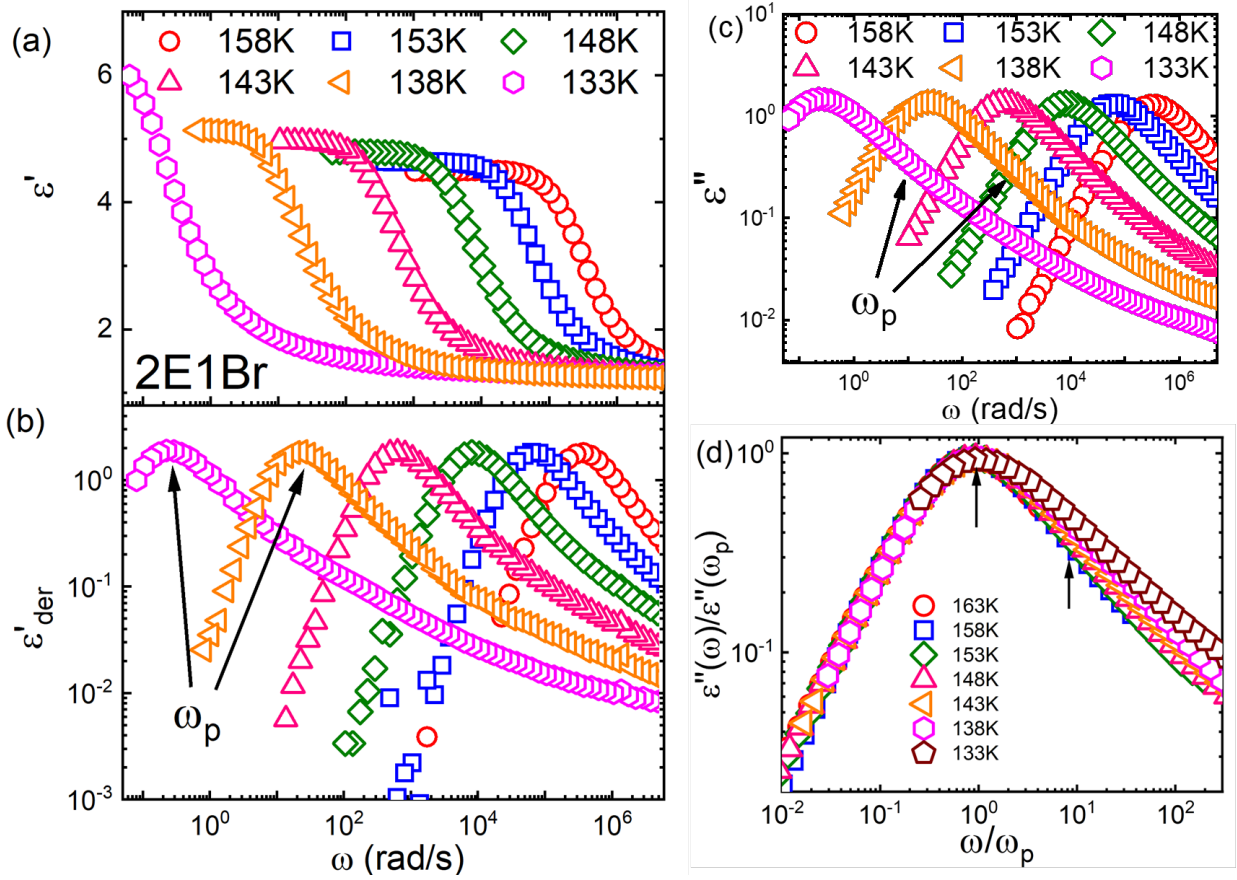

**Figure 2.** (a) Real part of the complex dielectric permittivity, $\varepsilon'(\omega)$. (b) Derivative spectra of $\varepsilon'(\omega)$, $\varepsilon'_{der}(\omega) = -\frac{\pi}{2}\frac{\partial \varepsilon'(\omega)}{\partial \ln\omega}$. (c) Imaginary part of the complex dielectric permittivity, $\varepsilon''(\omega)$. (d) Normalized dielectric loss permittivity, $\varepsilon''(\omega)/\varepsilon''(\omega_p)$ vs $\omega/\omega_p$, of 2E1Br, where $\omega_p$ and $\varepsilon''(\omega_p)$ are the corresponding characteristic frequencies and the dielectric amplitude of the main peak of $\varepsilon''(\omega)$.

We analyze the dielectric spectra in two different ways: Method (i) is based on a direct fit to the spectra with Havriliak-Negami (HN) functions:

$$\varepsilon^*(\omega) = \varepsilon'(\omega) - i\varepsilon''(\omega) = \varepsilon_\infty + \sum_k \frac{\Delta\varepsilon_k}{\left[1 + (i\omega\tau_{HN,k})^{\beta_k}\right]^{\gamma_k}} \quad (2)$$

where $\varepsilon^*(\omega)$ is the complex permittivity, $\tau_{HN,k}$, $\Delta\varepsilon_k$, $\beta_k$, and $\gamma_k$ are the characteristic HN time, the dielectric relaxation strength, the symmetric, and the asymmetric stretching parameter of the $k^{th}$ process, respectively. Method (ii) involves a relaxation time distribution analysis that can help to separate overlapping relaxation processes:

$$\varepsilon^*(\omega) = \varepsilon_\infty + \Delta\varepsilon \times \int_{-\infty}^{+\infty} \frac{g(ln\tau)}{1 + i\omega\tau} dln\tau \qquad (3)$$

Here $\Delta\varepsilon$ is the total dielectric relaxation strength of all relaxation processes, $\tau$ is the characteristic relaxation time, $g(ln\tau)$ is the relaxation time distribution density function and $\int_{-\infty}^{+\infty} g(ln\tau)dln\tau = 1$ holds. We apply a generalized regularization method[28] to solve the integral equation to obtain $g(ln\tau)$. Similar analyses have been discussed previously.[29-31] Ideally, $g(ln\tau)$ should have a sharp cut-off to zero beyond the terminal relaxation time. In practice, cut-off with finite slope can be achieved due to the limitation of any numerically analysis. At the same time, we would like to emphasize that the analysis provides one solution of the relaxation time distribution and can help separate the overlapping relaxation processes.

**Figures 3a and 3b** provide the representative HN function fit to $\varepsilon''(\omega)$ and $\varepsilon'_{der}(\omega)$ of 2E1Br at $T = 143\ K$ from the same set of HN function parameters. The dashed lines and the dash-dotted lines represent fits to Processes I and II, respectively. The solid lines are the sum of the two contributions. **Figure 3c** reflects the relaxation time distribution analysis, which yields two well-resolved relaxation processes: one is the characteristic relaxation time of Process I and the other of Process II. Thus, *both* the HN function fit and the relaxation time distribution analyses point to *two* distinct molecular relaxation processes of 2E1Br.

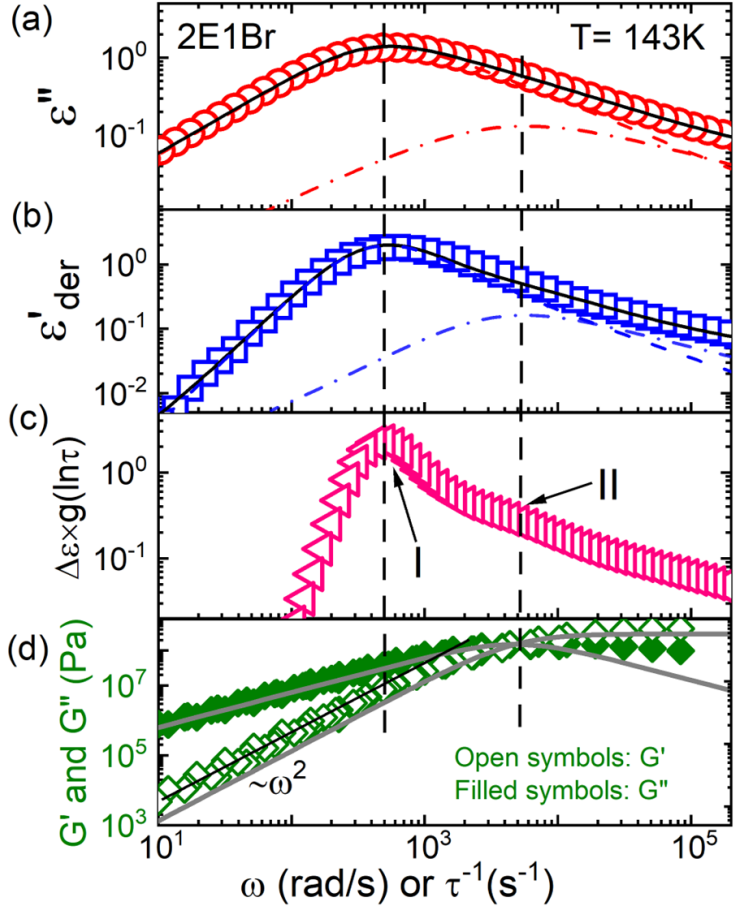

**Figure 3. (a)** Loss permittivity, $\varepsilon''(\omega)$ (red circles) and **(b)** derivative of the storage permittivity, $\varepsilon'_{der}(\omega)$ (blue squares) at $T = 143K$. The dashed and the dash-dotted lines represent HN fits to Process I and Process II, respectively. **(c)** Relaxation time distribution density function, $g(\ln\tau)$, at $T = 143K$. **(d)** Linear viscoelastic master curve obtained through time-temperature superposition at a reference temperature, $T = 143K$. The solid grey lines are the calculation from the single-mode Maxwell model, $G'(\omega) = \frac{G_0(\omega\tau)^2}{1+(\omega\tau)^2}$ and $G''(\omega) = \frac{G_0\omega\tau}{1+(\omega\tau)^2}$, with characteristic relaxation time of $\tau = \tau_\alpha^R$ and a plateau modulus $G_0 = 2.3\times10^8\ Pa$. The solid black line represents $G'(\omega) \sim \omega^2$, and the deviation between the experiments and the $G'(\omega) \sim \omega^2$ line defines the characteristic relaxation rate associated with the terminal flow, $\omega_f$. The terminal flow time is then $\tau_f = 1/\omega_f$.

To be more quantitative, we look into the parameters of the HN function fit (inset of **Figure 4a**): (i) The symmetric and the asymmetric stretching parameters of Process I are $\beta_I \approx 1.0$ and $\gamma_I \approx 0.7$ at all temperatures (inset of **Figure 4a**), signifying process I as a Debye-like process. (ii) The symmetric and the asymmetric stretching parameters of Process II are $\beta_{II} \approx 0.8$ and $\gamma_{II} \approx 0.6$ (inset of **Figure 4a**), which are close to the generic shape parameters for structural relaxations.[16]

(iii) The characteristic relaxation times of Process I, $\tau_I$, and of Process II, $\tau_{II}$, follow super-Arrhenius temperature dependences and can be described well by the Vogel-Fulcher-Tammann (VFT) relation[32-33]: $\log_{10}(\tau) = -A + \frac{B}{T-T_0}$, where $A$, $B$, and $T_0$ (Vogel temperature) are fit parameters (see **Table S1** of the **SM**). From the VFT fits (solid lines in **Figure 4a**), the dynamic glass transition temperatures of Process I and Process II at which $\tau = 100s$ are $T_{100,I} = 130\ K$ and $T_{100,II} = 128\ K$. The super-Arrhenius temperature dependences of Processes I and II suggest their cooperative nature. Another important feature is that the separation between $\tau_I$ and $\tau_{II}$ are larger and larger upon cooling, which was reflected as an increment in spectra broadening of $\varepsilon''(\omega)$ at high frequencies upon cooling **(Figure 2d)**. (iv) The dielectric relaxation amplitudes, $\Delta\varepsilon_I$ and $\Delta\varepsilon_{II}$ (**Figure 4b**), increase with cooling, which are not characteristics of a secondary relaxation.[34-35] (v) The Kirkwood-Fröhlich parameter,[25-27] $g_k$, is slightly larger than 1.0 and increases with cooling (inset of **Figure 4b,** also see **Table S2** of **SM**), implying the *presence* of parallel alignment of the molecular dipoles in 2E1Br. We note that previous BDS measurements of the 2E1Br did not discuss the presence of Process II and assigned the Process I as the structural relaxation for glass transition.[19-20, 36] These analyses thus raise the following important question: what are the molecular origins of these two processes? We answer this question through additional rheological measurements.

### 3.2 Rheology

The analyses from BDS measurements suggest the dynamics $T_g$ of the two processes, $T_{100,I} = 130\ K$ and $T_{100,II} = 128\ K$, which are too close to be differentiated from the $T_g = 131\ K$ from differential scanning calorimetry measurements (see **Figure S6** of **SM**). Linear rheology, on the other hand, can give an independent measure of both dynamics.[37-38] **Figure 3d** (or the enlarged **Figure S7a** of **SM**) provides the linear viscoelastic master curve of 2E1Br at a reference temperature $T = 143\ K$. The dynamic shift factors, $a_T$, are given in the inset of **Figure S7a**, from which one can identify the rheological structural relaxation time of 2E1Br at different temperatures, $\tau_\alpha^R$, where the superscript $R$ represents results from rheology. No vertical shifts, $b_T$, are needed for the construction of the master curve. A van Gurp-Palmen plot[39] has been provided in **Figure S7b** of the **SM** to demonstrate the rheological simplicity of 2E1Br and that time-temperature superposition holds at all testing temperatures. The peak of the loss modulus, $G''$, at high frequencies, $\omega_\alpha$, coincides with the crossover between $G'$ and $G''$, which is typical for van der Waals liquids and provides an estimate of $\tau_\alpha^R \approx \frac{1}{\omega_\alpha}$. In addition, the scaling behaviors of $G' \propto \omega^2$ and $G'' \propto \omega$ are observed at low frequencies, emphasizing the access of the terminal mode. We identify the terminal flow time, $\tau_f = \frac{1}{\omega_f}$, from the onset frequency, $\omega_f$, where $G' \propto \omega^2$ holds. $\omega_f$ is also the characteristic frequency the complex viscosity deviates from its zero-shear limit (see **Figure S7a** of the **SM**). Experimentally, $\frac{\tau_f}{\tau_\alpha^R} \approx 10$ is observed for 2E1Br, suggesting a noticeable separation between the structural relaxation and the terminal flow. Furthermore, the solid grey lines in **Figure 3d** provide a prediction of the single-mode Maxwell model on the linear viscoelastic properties with $\tau = \tau_\alpha^R$ and $G'(\omega) = \frac{G_0(\omega\tau)^2}{1+(\omega\tau)^2}, G''(\omega) = \frac{G_0\omega\tau}{1+(\omega\tau)^2}$ with $G_0$ the shear modulus at high frequencies.[40] A noticeable deviation between experiments and the Maxwell model has been observed at low frequencies, especially in $G'(\omega)$, further confirming the presence of separate dynamics slower than the structural relaxation of 2E1Br. Note that the Maxwell model describes well the viscoelastic spectra of non-polar van der Waals liquids in the absence of molecular associations.[41] Remarkably, $\tau_f$ and $\tau_\alpha^R$ (the filled symbols) agree excellently with $\tau_l$ and $\tau_{II}$ respectively (vertical lines in **Figure 3 and** open symbols in **Figure 4a**). The rheological analyses thus point to Process II as the structural relaxation and that Process I corresponds to the terminal relaxation of 2E1Br.

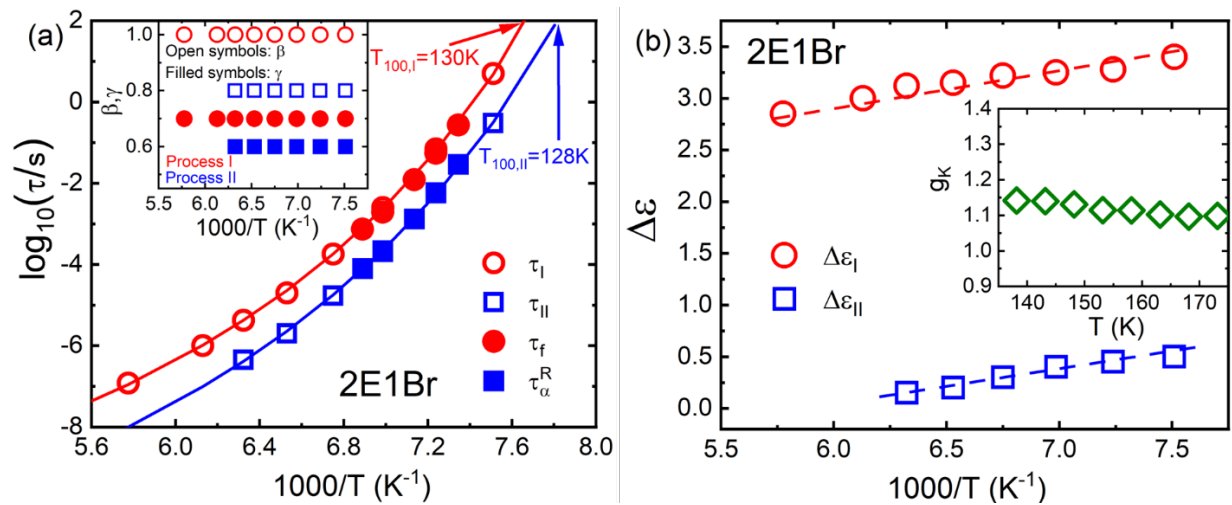

**Figure 4. (a)** Temperature dependence of Process I (empty red circles), and Process II (empty blue squares) from BDS and the temperature dependence of the $\tau_\alpha^R$ (filled blue diamonds) and $\tau_f$ (filled red circles) from rheology of 2E1Br. The inset presents the shape parameters of Process I and Process II from the HN function fit. **(b)** Temperature dependence of the dielectric relaxation strength from the HN fit for Process I (red circles), and Process II (blue squares). The inset represents the Kirkwood-Fröhlich factor, $g_k$ as a function of temperature.

We note that recent studies suggest the dielectric spectra of polar van der Waals liquids can be decomposed into a self-correlation part and the dipole-dipole cross-correlation part:[9, 16] The self-correlation part exhibits $\beta \approx 0.8$ and $\gamma \approx 0.6$ and the cross-correlation part is Debye-like. However, we do not believe the Debye-like Process I of 2E1Br is from the dipole-dipole cross-correlation due to the following: (i) Existing theoretical analyses[42] and recent simulations[14] suggest a dipole moment much larger than $1.56\,D$ is required to obtain the characteristic time associated with dipole-dipole cross-correlation 10 times slower than the self-part of the dipole-dipole correlation. (ii) The rheological measurements show the dynamics moduli associated with the terminal flow, $G'(\omega_f) \approx 8\times10^6$ Pa (see **Figure 3d**), is only ~3% of that of the glassy moduli, $G_0$, indicating the involvement of collective motion of a large amount of molecules for the terminal flow. On the other hand, the combined analyses of BDS and rheology suggest the formation of supramolecular chains of 2E1Br, which offers an explanation of the Process I and the observed $g_k > 1$ (see the inset of **Figure 4b**). The presence of supramolecular dynamics of 2E1Br might also explain the shoulder peak at wavevectors smaller than the first-sharp diffraction peak from previous small-angle x-ray scattering (SAXS) measurements (**Figure S8** of the **SM**).[20, 43] Note that neat alkanes do not have the pre-peak at low wavevector region smaller than the first-sharp diffraction peak.[44] Thus, we attribute supramolecular structure formation of 2E1Br to the finite dipole-dipole interactions of the C-Br group. Recent SAXS measurements also found a weak shoulder pre-peak or the absence of the pre-peak for heteroatom substitutes of oxygen in monohydroxy alcohols, such as the 2-ethyl-1-hexanethiol.[45] This observation does not conflict with the proposed relationship between the above discussion about the absence of the pre-peak since the C-S bond of 2-ethyl-1-hexanethiol has a dipole moment of $\mu \sim 0.7D$, which is much smaller than that of C-Br.[46]

### 3.3 The emergence of Debye-like process in other alkyl halides.

If the supramolecular structure formation of 2E1Br is due to the C-Br group, one should be able to observe similar supramolecular dynamics in other monofunctional halides. Hence, we explore further the dynamics of other monofunctional halides: 1-bromo-3,7-dimethyl octane (3,7D1OBr) and reanalyze the dielectric data from 1-chloro-2-ethyl hexane (2E1Cl). The dielectric data for 2E1Cl has already been published in Ref. 20. **Figure 5** presents the representative spectra of the dielectric storage permittivity, $\varepsilon'(\omega)$ (**Figure 5a**), the derivative spectra, $\varepsilon'_{der}(\omega)$ (**Figure 5b**), and the loss permittivity, $\varepsilon''(\omega)$ (**Figure 5c**), of 3,7D1OBr at temperatures $T =$ $138K, 143K, 148K, 153K, 158K$, and $163K$, where $\omega$ is the angular frequency. A dominant dielectric relaxation can be identified at each temperature from the step of $\varepsilon'(\omega)$ and from the characteristic peaks of both $\varepsilon'_{der}(\omega)$ and $\varepsilon''(\omega)$. A closer look into the $\varepsilon'_{der}(\omega)$ and $\varepsilon''(\omega)$ shows the presence of three dielectric relaxation peaks. Furthermore, normalizing $\varepsilon''(\omega)$ by the characteristic peak frequency, $\omega_p$, gives a nice overlapping in the dielectric response at the low-frequency region and an increment in spectral broadening at high frequencies upon cooling (**Figure 5d** for 3,7D1OBr and **Figure 6** for 2E1Cl). Similar dielectric broadening at high frequencies can be well-observed in the normalized dielectric derivative spectra $\varepsilon'_{der}(\omega)$ (**Figure S9** in the **SM**).

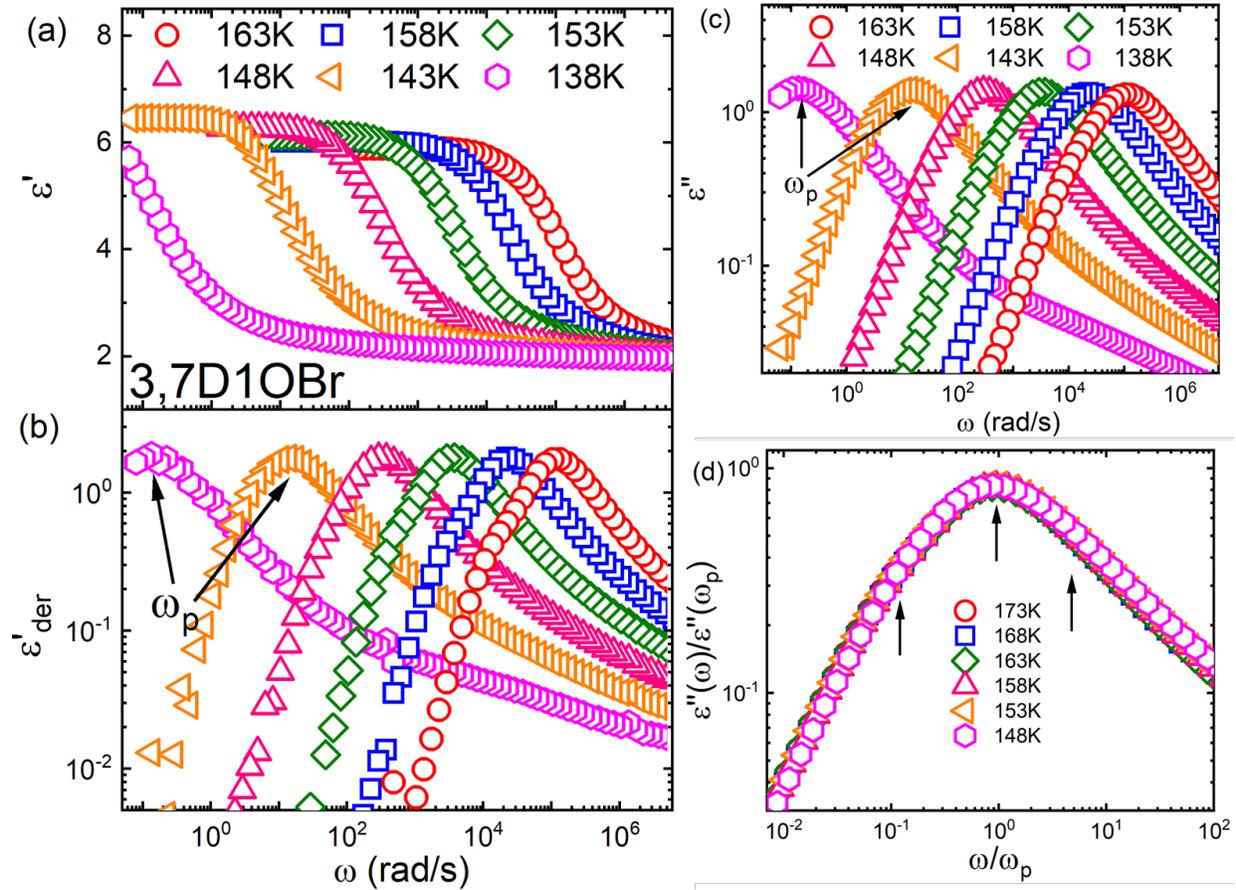

**Figure 5**. (a) Real part of the complex dielectric permittivity, $\varepsilon'(\omega)$. (b) Derivative spectra of $\varepsilon'(\omega)$, $\varepsilon'_{der}(\omega)$. (c) Imaginary part of the complex dielectric permittivity, $\varepsilon''(\omega)$. (d) Normalized imaginary

spectra, $\varepsilon''(\omega)/\varepsilon''(\omega_p)$ vs $\omega/\omega_p$, for 3,7D1OBr. The arrows indicate the locations of the processes from the normalized imaginary spectra.

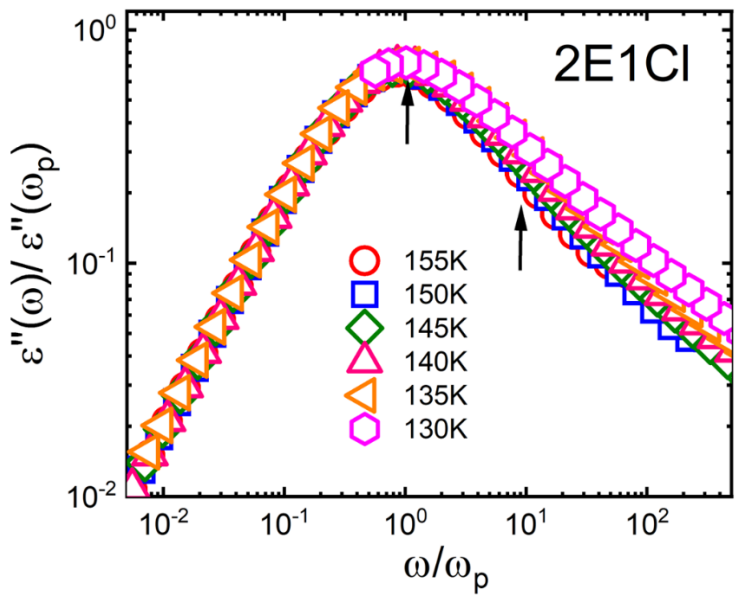

**Figure 6.** Normalized imaginary spectra, $\varepsilon''(\omega)/\varepsilon''(\omega_p)$ vs $\omega/\omega_p$, for 2E1Cl. Data is a reanalysis of dielectric data in Ref. 20. The arrows indicate the locations of the processes from the normalized imaginary spectra.

**Figure 7a** and **Figure 7b** provide the representative analyses of the dielectric spectra of 3,7D1OBr and 2E1Cl at $T = 148\ K$ and $T = 135\ K$ respectively. Interestingly, three relaxation processes are found in 3,7D1OBr, and two are observed for 2E1Cl. The dashed lines, the dotted lines, and the dash-dotted lines are the corresponding HN functions fit to 3,7D1OBr, and the dashed lines and the dashed-dotted lines are the corresponding HN functions fit to 2E1Cl, as shown in **Figures 7a** and **7b**. In addition, we have conducted the relaxation time distribution analyses to resolve the overlapping processes. At the same time, linear viscoelastic master curves of 3,7D1OBr and 2E1Cl at reference temperatures the same as the BDS measurements have been included for direct comparison. In analogy to 2E1Br, the molecular processes slower than the structural relaxation point to the presence of supramolecular dynamics.

We follow the same definition as for 2E1Br to assign the processes from low frequencies to high frequencies as Processes I (the dashed lines), II* (the dotted lines), and II (the dash-dotted lines) for 3,7D1OBr or Processes I (the dashed lines) and II (the dash-dotted lines) for 2E1Cl. The following interesting features are worth noting: (i) The characteristic times of Process II of 2E1Cl and 3,7D1OBr agree well with the structural relaxation from the corresponding rheology measurements (**Figures 7a and 7b**). The enlarged linear viscoelastic master curves of 3,7D1OBr and the 2E1Cl are given in **Figures S10a** and **S10b** of **SM** respectively. A van Gurp-Palmen plot[39] has been provided in **Figure S11a and S11b** in the **SM** respectively to demonstrate the rheological simplicity of 3,7D1OBr and 2E1Cl, and the hold of the time-temperature superposition principles. (ii) Similar to 2E1Br, the dielectric amplitudes of Process II of 2E1Cl and 3,7D1OBr increase with cooling, and their symmetric and asymmetric stretching parameters are $\beta \approx 0.7$ and $\gamma \approx 0.6$ (**Figures 8c and 8d**), which are close to the recently revealed generic features of the dielectric dispersion of molecular liquids.[16] These features suggest Process II of 2E1Cl and 3,7D1OBr the structural relaxation process. (iii) The Process I of 3,7D1OBr is a Debye process with $\beta_I = 1.0$ and $\gamma_I = 1.0$, and that of 2E1Cl is Debye-like ($\beta_I = 1.0$ and $\gamma_I = 0.75$) (insets of **Figures 8c** and **8d**). The characteristic times of the Debye or Debye-like process of 3,7D1OBr and 2E1Cl match well with their terminal relaxation time for flow (**Figures 8a** and **8b** or **Figures S10a** and **S10b**), which might not have the same temperature dependence from the structural relaxation. Similar features have been observed for the relationship between Debye relaxation and the structural relaxation of monohydroxy alcohols.[17-18, 41, 47] Furthermore, albeit 3,7D1OBr and 2E1Br have the identical C-Br polar moiety, the separation between terminal flow and the structural relaxation of 3,7D1OBr, $\tau_I/\tau_{II} \approx 40$, is noticeably larger than that of 2E1Br. This further suggests the Debye Process I of 2E1Br is not due to the recently proposed dipole-dipole cross-correlation in polar molecular liquids.[16, 48] (iv) For 3,7D1OBr, a process (Process II*) shows up in the intermediate frequency between the structural relaxation and the Debye process, in analogy to that of monohydroxy alcohols having ring-like supramolecular structures, such as 4-methyl-3-heptanol (4M3H).[49-50] The origin of this intermediate process is not entirely clear at this moment. We note that 3,7-dimethyl-1-octanol (3,7D1O) has an strong Debye process and has often been viewed as a chain former.[51-52] Due to the much weaker intermolecular interactions between C-Br group of 3,7D1OBr as compared to the stronger hydrogen bonding of 3,7D1O, a coexistence of chain-like structures and ring-like structures might take place, where the ring-like structures could give the Process II*. Nevertheless, these observations demonstrate the presence of supramolecular relaxation in other monofunctional halides, despite their weak polar-polar intermolecular interactions.

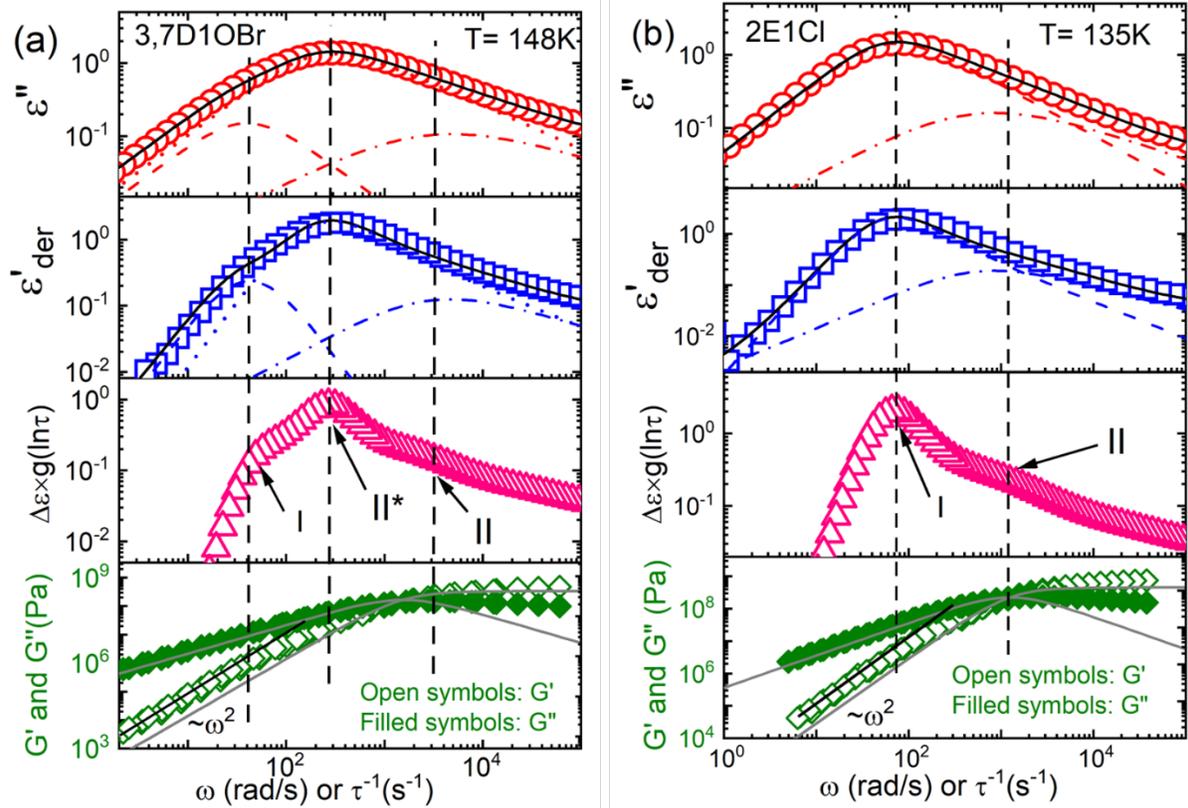

**Figure 7.** $\varepsilon''(\omega)$ (red circles), $\varepsilon'_{der}(\omega)$ (blue squares), $g(ln\tau)$, and linear viscoelastic master curves of (a) 37D1OBr at T=148K, and (b) 2E1Cl at T=135K. The dashed (Process I), dot (Process II$^*$) and dash-dotted (Process II) lines of **Figure 7a** represent the HN function fit to the dielectric spectra of 3,7D1OBr. The dashed (Process I) and dash-dotted (Process II) lines of **Figure 7b** give the HN function fit to the dielectric spectra of 2E1Cl. The solid black lines in the $\varepsilon''(\omega)$ and $\varepsilon'_{der}(\omega)$ panels are the sum of the HN functions fit. The grey solid lines of the rheology panels are the single-mode Maxwell model prediction with characteristic relaxation time of $\tau = \tau_\alpha^R$ and $G_0 = 3.0 \times 10^8\ Pa$ for 3,7D1OBr and $\tau = \tau_\alpha^R$ and $G_0 = 5.0 \times 10^8\ Pa$ for 2E1Cl. The solid black line of the rheology panels represents $G'(\omega) \propto \omega^2$. The terminal flow time is defined as $\tau_f = 1/\omega_f$, where $G'(\omega) \propto \omega^2$ holds at $\omega \leq \omega_f$.

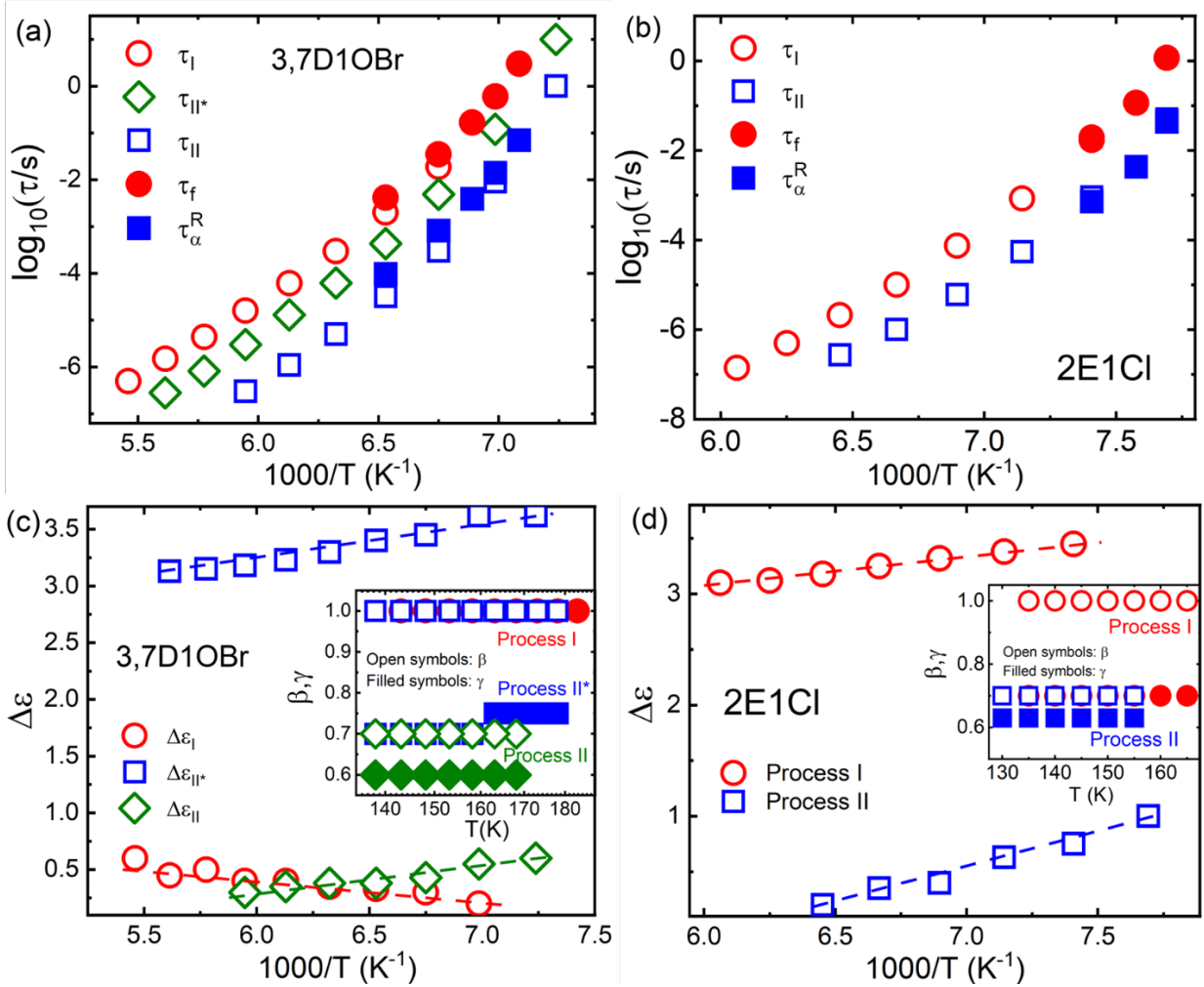

**Figure 8. (a)** Temperature dependence of $\tau_I$ (open red circles), $\tau_{II^*}$ (open olive diamonds), and $\tau_{II}$ (open blue squares), as well as the temperature dependence of the $\tau_\alpha^R$ (filled blue squares) and $\tau_f$ (filled red circles) of 3,7D1OBr. **(b)** Temperature dependence of $\tau_I$ (open red circles) and $\tau_{II}$ (open blue squares) from BDS and the temperature dependence of the $\tau_\alpha^R$ (filled blue squares) and $\tau_f$ (filled red circles) of 2E1Cl. **(c)** Dielectric relaxation strength, $\Delta\varepsilon$, of 3,7D1OBr for Process I (red circles), Process II$^*$ (blue squares), and Process II (olive diamonds). **(d)** Dielectric relaxation strength, $\Delta\varepsilon$, of 2E1Cl of Process I (red circles) and Process II (blue squares). The insets of the panels **(c)** and **(d)** give the corresponding shape parameters of the HN functions, $\beta$ (open symbols) and $\gamma$ (filled symbols)

### 3.4 The origin of the Debye-like Process I

To understand the Process I in the presently studied liquids, we focus on the two following observations: (i) the Process I is Debye-like and is dielectrically active, and (ii) the characteristic time of Process I is comparable with that of the terminal flow. These features suggest the presence of supramolecular chain structures of these mono-functional halides. At the same time, the mono-functional halides exhibiting supramolecular dynamics coupling with the terminal flow. Similar observations were found in monohydroxy alcohols (MAs) displaying supramolecular chain structures.[17-18, 41, 47] Compared with MAs that comprise chain structures, such as 2E1H (2-ethyl-1-hexanol) or 5M2H (5-methyl-2-hexanol), the time scale separation between the Debye-like process (Process I) and the structural relaxation (Process II) are much smaller (on the order of 10 for 2E1Br and 2E1Cl or 40 for 3,7D1OBr). This indicates their much smaller supramolecular chain sizes than 2E1H or 5M2H. According to the recently proposed living polymer model,[17-18] a supramolecular chain length, $N \approx \tau_f/\tau_\alpha$ is found at $\tau_\alpha \geq \tau_B$, implying the presence of supramolecular structures of 10-40 molecules in these monofunctional halides. We would like to emphasize that living polymer model emphasizes the collective dynamics of supramolecular chain structures for terminal relaxation, which is different from the sizes of the cooperatively rearranging region (CRR) for structural relaxation.[53-55] At the same time, the supramolecular chain structures formation should affect the structural relaxation, although the relationship between the supramolecular structures formation and the sizes of CRR remains unclear at this moment. If we assume a Gaussian distribution of the supramolecular structures and the characteristic ratio of $C_\infty \approx 7.0$, the supramolecular chain of ~10 molecules of 2E1Br has a characteristic size (i.e. radius of gyration) of 0.7 nm, which agrees well with its cluster sizes estimated from the small-angle x-ray scattering measurements (see **Figure S8** of the **SM**).[20] Moreover, the temperature dependence of the supramolecular chain sizes can offer an estimate of the enthalpy of the association, $N(T) \sim \tau_f/\tau_\alpha \sim K_{eq}^{1/2} \approx \exp(E_a/(2RT))$, with $K_{eq}$ being the equilibrium reaction rate constant for molecular association and dissociation, $E_a$ the enthalpy of association/dissociation reaction, and $R$ the gas constant.[17-18] **Figure 9** provides the temperature dependence of $N(T)$ from dynamics measurements, which exhibit Arrhenius temperature dependence with apparent activation energies of ~2.6 kJ/mol for 2E1Br, ~7.7 kJ/mol for 2E1Cl, and ~1.7 kJ/mol for 3,7D1OBr. As a result, the enthalpies of the association/dissociation reaction are $E_a \approx 5.2\ kJ/mol$ for 2E1Br, $E_a \approx 15.4\ kJ/mol$ for 2E1Cl, and $E_a \approx 3.4\ kJ/mol$ for 3,7D1OBr. Interestingly, the estimated enthalpy change agrees well with the dipole-dipole interaction strength with 2E1Cl > 2E1Br ~ 3,7D1OBr.

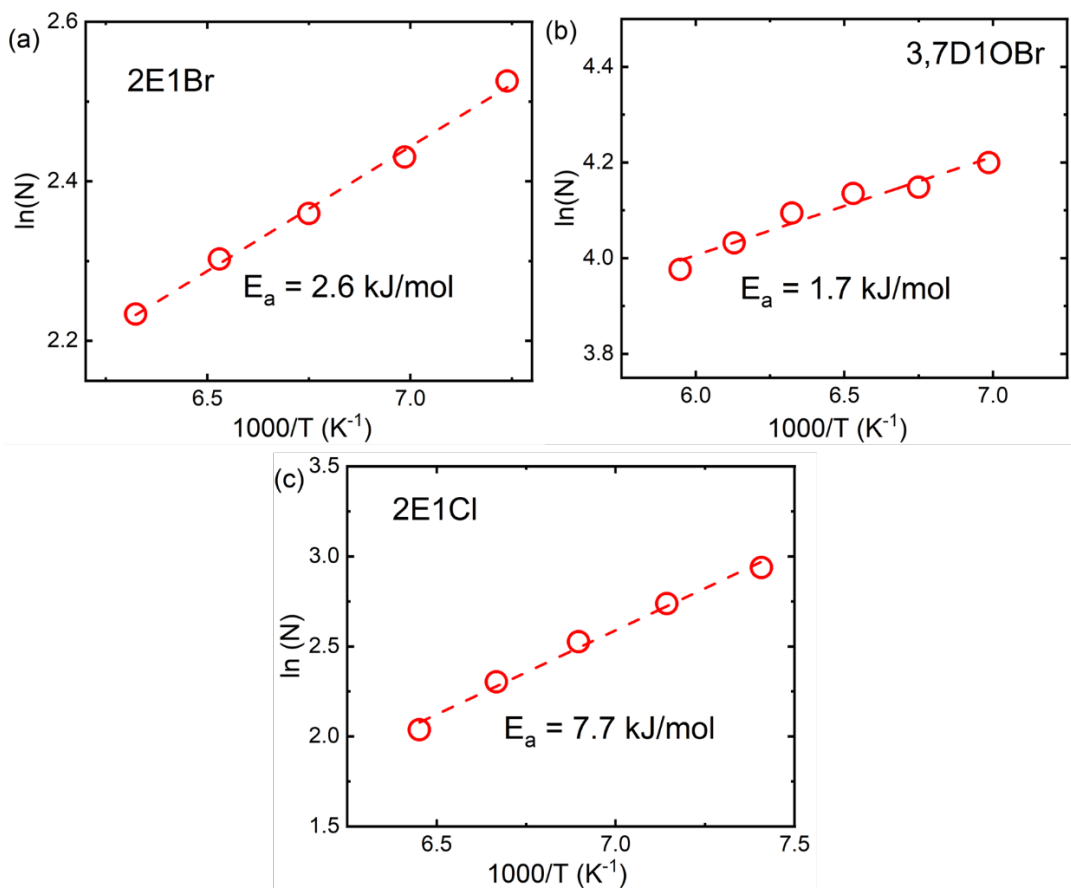

**Figure 9.** Temperature dependence of the supramolecular chain size, $N \sim \tau_f/\tau_\alpha$ for **(a)** 2E1Br, **(b)** 3,7D1OBr and **(c)** 2E1Cl.

Lastly, we would like to comment on the formation of the large supramolecular chain-like structures in these molecular liquids. For 2E1Br, 2E1Cl, or 3,7D1OBr, the maximum dipole-dipole interaction is when the dipoles orient parallel with each other, i.e. the chain conformation. Assuming $\mu \approx 1.5\ D$, the relative dielectric constant $\varepsilon_r \approx 5.0$ (**Figure 1**), vacuum permittivity $\varepsilon_0 \approx 8.854 \times 10^{-12} F/m$, and distance $r \approx 2-3$ Å, the interaction potential energy of two neighboring dipole aligning parallel to each other is[56] $U \approx \frac{\mu^2}{2\pi\varepsilon_r\varepsilon_0 r^3} \approx 3-10 \times 10^{-21}\ J$, which is equivalent to $1.8\ kJ/mol\ (\sim 1.5\ k_B T$ at $T = 143\ K)$ to $6.02\ kJ/mol\ (\sim 5.0\ k_B T$ at $T = 143\ K)$. Thus, the dipole-dipole interaction of the C-Cl or C-Br groups can overcome the thermal energy at low temperatures and lead to molecular association. Remarkably, the above estimated dipole-dipole interactions agree reasonably well with the enthalpy ($3.4\ kJ/mol$ to $15.4\ kJ/mol$) of the molecular association/dissociation, further confirming the presence of supramolecular chain structures. At the same time, we do believe other types of supramolecular structures co-exist in these liquids along with the supramolecular chain structures, such as micellar structures or supramolecular ring structures. This can be indirectly inferred through the presence of multiple relaxation process and a moderate $g_k > 1$ values. For instance, the emergence of the Process II* of 3,7D1OBr might correlate with the non-chain-like structures. From this perspective, the combined BDS and rheology measurements allow one to characterize the supramolecular chain structures, while additional studies are needed to gain an in-depth characterization of the non-chain-like structures of these polar liquids.

## 4. Conclusion

In conclusion, we have performed dielectric and rheological measurements to investigate the dynamics of a group of mono-functional halides, the 2E1Cl, 2E1Br, and 3,7D1OBr. BDS measurements demonstrated (i) clear signatures of multiple molecular relaxation processes with super-Arrhenius temperature dependence, and (ii) the emergence of slow dynamics (more than 10 times slower) compared with their structural relaxation. At the same time, linear rheology revealed large separations between the structural relaxation and terminal flow. A detailed comparison of the BDS and rheology reveals: (i) the rheological structural relaxation matches with the high-frequency BDS process (Process II), and (ii) the rheological terminal relaxation agrees with the low-frequency BDS process (the Debye-like Process I). We explain the emergence of the Process I (the slow Debye-like process) through the formation of transient supramolecular chain structures in these polar van der Waals liquids. The results suggest that weak van der Waals interactions could induce large supramolecular structures and collective slow supramolecular dynamics in the deep supercooled region. Since supramolecular structures affect strongly many fundamental macroscopic properties of liquids, including the dielectric constant,[3] the glass formation,[57] and the viscosity of molecular liquids,[7, 58] we anticipate the results to impact strongly the understanding of fundamental properties of polar liquids.

**Acknowledgment**

This work was supported by the Michigan State University Discretionary Funding Initiative (MSU-DFI). The authors would like to thank Sean Walsh and Prof. Robert Malezcka's group from the Department of Chemistry at Michigan State University for their help on the purification and characterization of 2E1Br and 3,7D1OBr. We thank Dr. Shinian Cheng at University of Tennessee to help on the DSC measurements of 2E1Br.

# References

1. Hansen, J. P.; McDonald, I. R., *Theory of Simple Liquids*. Academic Press: 1986.
2. Wei, D.; Patey, G. N., Rotational motion in molecular liquids. *J. Chem. Phys.* **1989,** *91* (11), 7113-7129.
3. Zhuang, B.; Wang, Z.-G., Statistical field theory for polar fluids. *J. Chem. Phys.* **2018,** *149* (12), 124108.
4. Teixeira, P. I. C.; Tavares, J. M.; Telo da Gama, M. M., The effect of dipolar forces on the structure and thermodynamics of classical fluids. *J. Phys.: Condens. Matter* **2000,** *12* (33), R411.
5. Samanta, S.; Kim, S.; Saito, T.; Sokolov, A. P., Polymers with Dynamic Bonds: Adaptive Functional Materials for a Sustainable Future. *J. Phys. Chem. B* **2021,** *125* (33), 9389-9401.
6. Zhang, Z.; Chen, Q.; Colby, R. H., Dynamics of associative polymers. *Soft Matter* **2018,** *14* (16), 2961-2977.
7. Ediger, M. D.; Angell, C. A.; Nagel, S. R., Supercooled Liquids and Glasses. *J. Phys. Chem.* **1996,** *100* (31), 13200-13212.
8. Pabst, F.; Helbling, A.; Gabriel, J.; Weigl, P.; Blochowicz, T., Dipole-dipole correlations and the Debye process in the dielectric response of nonassociating glass forming liquids. *Phys. Rev. E* **2020,** *102* (1), 010606.
9. Böhmer, T.; Pabst, F.; Gabriel, J. P.; Zeißler, R.; Blochowicz, T., On the spectral shape of the structural relaxation in supercooled liquids. *J. Chem. Phys.* **2025,** *162* (12), 120902.
10. Arrese-Igor, S.; Alegría, A.; Colmenero, J., Non-simple flow behavior in a polar van der Waals liquid: Structural relaxation under scrutiny. *J. Chem. Phys.* **2023,** *158* (17), 174504.
11. Tarnacka, M.; Czaderna-Lekka, A.; Wojnarowska, Z.; Kamiński, K.; Paluch, M., Nature of Dielectric Response of Phenyl Alcohols. *J. Phys. Chem. B* **2023,** *127* (27), 6191-6196.
12. Wang, Y.; Griffin, P. J.; Holt, A.; Fan, F.; Sokolov, A. P., Observation of the slow, Debye-like relaxation in hydrogen-bonded liquids by dynamic light scattering. *J. Chem. Phys.* **2014,** *140* (10), 104510.
13. Koperwas, K.; Gapiński, J.; Wojnarowska, Z.; Patkowski, A.; Paluch, M., Experimental examination of dipole-dipole cross-correlations by dielectric spectroscopy, depolarized dynamic light scattering, and computer simulations of molecular dynamics. *Phys. Rev. E* **2024,** *109* (3), 034608.
14. Koperwas, K.; Paluch, M., Computational Evidence for the Crucial Role of Dipole Cross-Correlations in Polar Glass-Forming Liquids. *Phys. Rev. Lett.* **2022,** *129* (2), 025501.
15. Becher, M.; Lichtinger, A.; Minikejew, R.; Vogel, M.; Rössler, E. A., NMR Relaxometry Accessing the Relaxation Spectrum in Molecular Glass Formers. *Int. J. Mol. Sci.* **2022,** *23* (9), 5118.
16. Pabst, F.; Gabriel, J. P.; Böhmer, T.; Weigl, P.; Helbling, A.; Richter, T.; Zourchang, P.; Walther, T.; Blochowicz, T., Generic Structural Relaxation in Supercooled Liquids. *J. Phys. Chem. Lett.* **2021,** *12* (14), 3685-3690.
17. Cheng, S.; Patil, S.; Cheng, S., Hydrogen Bonding Exchange and Supramolecular Dynamics of Monohydroxy Alcohols. *Phys. Rev. Lett.* **2024,** *132* (5), 058201.
18. Patil, S.; Sun, R.; Cheng, S.; Cheng, S., Molecular Mechanism of the Debye Relaxation in Monohydroxy Alcohols Revealed from Rheo-Dielectric Spectroscopy. *Phys. Rev. Lett.* **2023,** *130* (9), 098201.

19. Preuß, M.; Gainaru, C.; Hecksher, T.; Bauer, S.; Dyre, J.; Richert, R.; Böhmer, R., Experimental studies of Debye-like process and structural relaxation in mixtures of 2-ethyl-1-hexanol and 2-ethyl-1-hexyl bromide. *J. Chem. Phys.* **2012,** *137* (14), 144502.
20. Büning, T.; Lueg, J.; Bolle, J.; Sternemann, C.; Gainaru, C.; Tolan, M.; Böhmer, R., Connecting structurally and dynamically detected signatures of supramolecular Debye liquids. *J. Chem. Phys.* **2017,** *147* (23), 234501.
21. Bauer, S.; Moch, K.; Münzner, P.; Schildmann, S.; Gainaru, C.; Böhmer, R., Mixed Debye-type liquids studied by dielectric, shear mechanical, nuclear magnetic resonance, and near-infrared spectroscopy. *J. Non-Cryst. Solids* **2015,** *407*, 384-391.
22. Böhmer, R.; Gainaru, C.; Richert, R., Structure and dynamics of monohydroxy alcohols—Milestones towards their microscopic understanding, 100 years after Debye. *Phys. Rep.* **2014,** *545* (4), 125-195.
23. Gainaru, C.; Meier, R.; Schildmann, S.; Lederle, C.; Hiller, W.; Rössler, E. A.; Böhmer, R., Nuclear-Magnetic-Resonance Measurements Reveal the Origin of the Debye Process in Monohydroxy Alcohols. *Phys. Rev. Lett.* **2010,** *105* (25), 258303.
24. Smyth, C. P., *Dielectric Behavior and Structure: Dielectric Constant and Loss, Dipole Moment and Molecular Structure*. McGraw-Hill: 1955.
25. Singh, L. P.; Alba-Simionesco, C.; Richert, R., Dynamics of glass-forming liquids. XVII. Dielectric relaxation and intermolecular association in a series of isomeric octyl alcohols. *J. Chem. Phys.* **2013,** *139* (14), 144503.
26. Fröhlich, H., *Theory of dielectrics : dielectric constant and dielectric loss*. Clarendon Press: 1949.
27. Kirkwood, J. G., The Dielectric Polarization of Polar Liquids. *J. Chem. Phys.* **1939,** *7* (10), 911-919.
28. Roths, T.; Marth, M.; Weese, J.; Honerkamp, J., A generalized regularization method for nonlinear ill-posed problems enhanced for nonlinear regularization terms. *Computer Physics Communications* **2001,** *139* (3), 279-296.
29. Carroll, B.; Cheng, S.; Sokolov, A. P., Analyzing the Interfacial Layer Properties in Polymer Nanocomposites by Broadband Dielectric Spectroscopy. *Macromolecules* **2017,** *50* (16), 6149-6163.
30. Cheng, S.; Mirigian, S.; Carrillo, J.-M. Y.; Bocharova, V.; Sumpter, B. G.; Schweizer, K. S.; Sokolov, A. P., Revealing spatially heterogeneous relaxation in a model nanocomposite. *J. Chem. Phys.* **2015,** *143* (19), 194704.
31. Cheng, S.; Holt, A. P.; Wang, H.; Fan, F.; Bocharova, V.; Martin, H.; Etampawala, T.; White, B. T.; Saito, T.; Kang, N.-G.; Dadmun, M. D.; Mays, J. W.; Sokolov, A. P., Unexpected Molecular Weight Effect in Polymer Nanocomposites. *Phys. Rev. Lett.* **2016,** *116* (3), 038302.
32. Tammann, G.; Hesse, W., Die Abhängigkeit der Viscosität von der Temperatur bie unterkühlten Flüssigkeiten. *Zeitschrift für anorganische und allgemeine Chemie* **1926,** *156* (1), 245-257.
33. Fulcher, G., Analysis of Recent Measurement of the Viscosity of Glasses. *J. Am. Ceram. Soc.* **2006,** *8*, 339-355.
34. Popov, I.; Cheng, S.; Sokolov, A. P., Broadband Dielectric Spectroscopy and Its Application in Polymeric Materials. In *Macromolecular Engineering*, 2022; pp 1-39.
35. Kremer, F.; Schönhals, A., *Broadband Dielectric Spectroscopy*. Springer Berlin Heidelberg: 2002.
25

36. Lederle, C.; Hiller, W.; Gainaru, C.; Böhmer, R., Diluting the hydrogen bonds in viscous solutions of n-butanol with n-bromobutane: II. A comparison of rotational and translational motions. *J. Chem. Phys.* **2011,** *134* (06), 064512.
37. Cheng, S.; Xie, S.-J.; Carrillo, J.-M. Y.; Carroll, B.; Martin, H.; Cao, P.-F.; Dadmun, M. D.; Sumpter, B. G.; Novikov, V. N.; Schweizer, K. S.; Sokolov, A. P., Big Effect of Small Nanoparticles: A Shift in Paradigm for Polymer Nanocomposites. *ACS Nano* **2017,** *11* (1), 752-759.
38. Yang, J.; Melton, M.; Sun, R.; Yang, W.; Cheng, S., Decoupling the Polymer Dynamics and the Nanoparticle Network Dynamics of Polymer Nanocomposites through Dielectric Spectroscopy and Rheology. *Macromolecules* **2020,** *53* (1), 302-311.
39. Trinkle, S.; Friedrich, C., Van Gurp-Palmen-plot: a way to characterize polydispersity of linear polymers. *Rheo. Acta* **2001,** *40* (4), 322-328.
40. Ferry, J. D., *Viscoelastic properties of polymers*. Wiley, New York: 1980.
41. Gainaru, C.; Figuli, R.; Hecksher, T.; Jakobsen, B.; Dyre, J. C.; Wilhelm, M.; Böhmer, R., Shear-Modulus Investigations of Monohydroxy Alcohols: Evidence for a Short-Chain-Polymer Rheological Response. *Phys. Rev. Lett.* **2014,** *112* (9), 098301.
42. Déjardin, P.-M.; Titov, S. V.; Cornaton, Y., Linear complex susceptibility of long-range interacting dipoles with thermal agitation and weak external ac fields. *Phys. Rev. B* **2019,** *99* (2), 024304.
43. Bierwirth, S. P.; Bolle, J.; Bauer, S.; Sternemann, C.; Gainaru, C.; Tolan, M.; Böhmer, R., Scaling of Suprastructure and Dynamics in Pure and Mixed Debye Liquids. In *The Scaling of Relaxation Processes*, Kremer, F.; Loidl, A., Eds. Springer International Publishing: Cham, 2018; pp 121-171.
44. Franks, N. P.; Abraham, M. H.; Lieb, W. R., Molecular Organization of Liquid n-Octanol: An X-ray Diffraction Analysis. *J. Pharm. Sci.* **1993,** *82* (5), 466-470.
45. Jurkiewicz, K.; Hachuła, B.; Kamińska, E.; Grzybowska, K.; Pawlus, S.; Wrzalik, R.; Kamiński, K.; Paluch, M., Relationship between Nanoscale Supramolecular Structure, Effectiveness of Hydrogen Bonds, and Appearance of Debye Process. *J. Phys. Chem. C.* **2020,** *124* (4), 2672-2679.
46. Harrison, J. F., Relationship between the Charge Distribution and Dipole Moment Functions of CO and the Related Molecules CS, SiO, and SiS. *J. Phys. Chem. A* **2006,** *110* (37), 10848-10857.
47. Singh, L. P.; Richert, R., Watching Hydrogen-Bonded Structures in an Alcohol Convert from Rings to Chains. *Phys. Rev. Lett.* **2012,** *109* (16), 167802.
48. Böhmer, T.; Pabst, F.; Gabriel, J. P.; Zeißler, R.; Blochowicz, T., On the spectral shape of the structural relaxation in supercooled liquids. *J. Chem. Phys.* **2025,** *162* (12),120902.
49. Hecksher, T.; Jakobsen, B., Communication: Supramolecular structures in monohydroxy alcohols: Insights from shear-mechanical studies of a systematic series of octanol structural isomers. *J. Chem. Phys.* **2014,** *141* (10), 101104.
50. Böhmer, T.; Richter, T.; Gabriel, J. P.; Zeißler, R.; Weigl, P.; Pabst, F.; Blochowicz, T., Revealing complex relaxation behavior of monohydroxy alcohols in a series of octanol isomers. *J. Chem. Phys.* **2023,** *159* (05), 054501.
51. Xu, D.; Feng, S.; Wang, J.-Q.; Wang, L.-M.; Richert, R., Entropic Nature of the Debye Relaxation in Glass-Forming Monoalcohols. *J. Phys. Chem. Lett* **2020,** *11* (14), 5792-5797.

52. Bauer, S.; Burlafinger, K.; Gainaru, C.; Lunkenheimer, P.; Hiller, W.; Loidl, A.; Böhmer, R., Debye relaxation and 250 K anomaly in glass forming monohydroxy alcohols. *J. Chem. Phys.* **2013,** *138* (9), 094505.
53. Albert, S.; Bauer, T.; Michl, M.; Biroli, G.; Bouchaud, J.-P.; Loidl, A.; Lunkenheimer, P.; Tourbot, R.; Wiertel-Gasquet, C.; Ladieu, F., Fifth-order susceptibility unveils growth of thermodynamic amorphous order in glass-formers. *Science* **2016,** *352* (6291), 1308-1311.
54. Cheng, S.; Sokolov, A. P., Correlation between the temperature evolution of the interfacial region and the growing dynamic cooperativity length scale. *J. Chem. Phys.* **2020,** *152* (9), 094904.
55. Bauer, T.; Lunkenheimer, P.; Loidl, A., Cooperativity and the Freezing of Molecular Motion at the Glass Transition. *Phys. Rev. Lett.* **2013,** *111* (22), 225702.
56. Riande, E.; Diaz-Calleja, R., *Electrical properties of polymers*. Marcel Dekker, Inc: New York, 2004.
57. Paluch, M.; Knapik, J.; Wojnarowska, Z.; Grzybowski, A.; Ngai, K. L., Universal Behavior of Dielectric Responses of Glass Formers: Role of Dipole-Dipole Interactions. *Phys. Rev. Lett.* **2016,** *116* (2), 025702.
58. Martinez, L. M.; Angell, C. A., A thermodynamic connection to the fragility of glass-forming liquids. *Nature* **2001,** *410* (6829), 663-667.

**For Table of Contents only (TOC):**

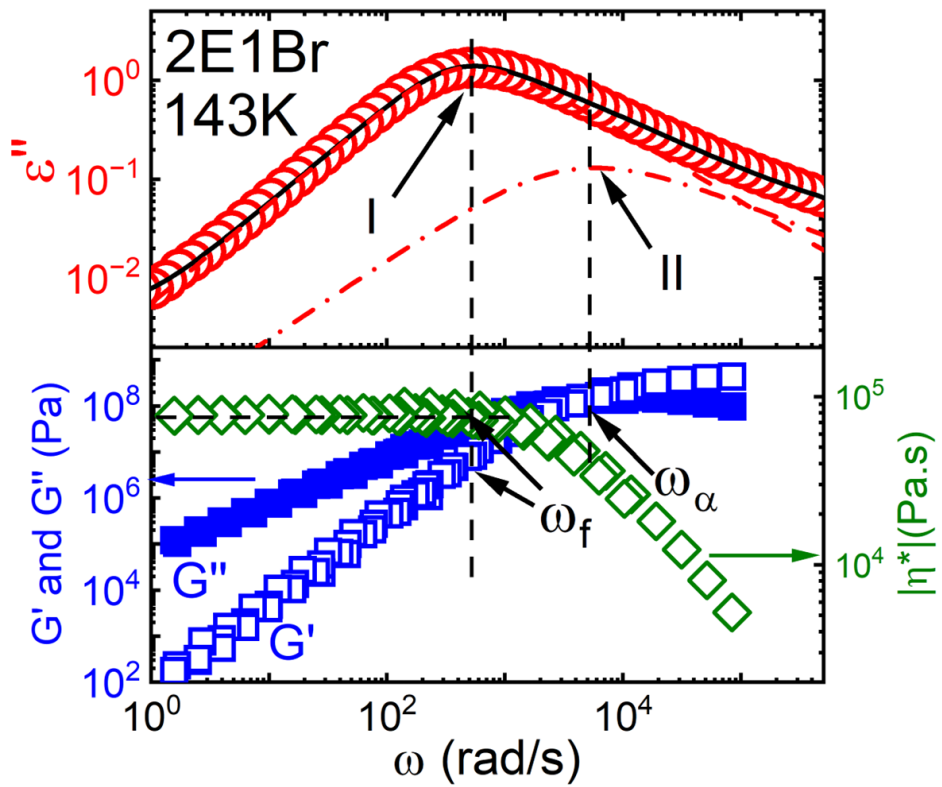

# Evidence for supramolecular dynamics of non-hydrogen bonding polar van der Waals liquids

Shalin Patil,[1] Catalin Gainaru,[2] Roland Böhmer,[3] and Shiwang Cheng[1]

[1] *Department of Chemical Engineering and Materials Science, Michigan State University, East Lansing, MI 48824, United States*

[2] *Chemical Sciences Division, Oak Ridge National Laboratory, Oak Ridge, Tennessee, 37831, United States*

[3] *Fakultät für Physik, Technische Universität Dortmund, Dortmund 44227, Germany*

Correspondence should be addressed to Shiwang Cheng at chengsh9@msu.edu

## 1 Methods

### 1.1 Distillation of 2E1Br and 3,7D1OBr

The 2E1Br and 3,7D1OBr purchased from Sigma-Aldrich contains impurities. A vacuum distillation was performed to purify them. In particular, around 10 ml of 2E1Br (or 3,7D1OBr) was taken in a 100 ml round bottom flask under heating and vacuum. The temperature of the 2E1Br (or 3,7D1OBr) in the round bottom flask was maintained at 338 K (403 K for 3,7D1OBr) through an oil bath. A vacuum of 21 mbar (2,100 Pa) was applied to facilitate the distillation process. The round bottom flask was connected to a distillation head with a thermometer adapter. A Liebig condenser was attached to the distillation head, and the condenser outlet was connected to three receiving flasks via an adapter. All joints were secured using with keck clips to ensure stability. The receiving flasks were rotated after collecting the initial distillate which contained mostly cyclohexane (as it is the most volatile component). The distilled fraction of 2E1Br (or 3,7D1OBr) was then collected and used for further characterizations and measurements.

### 1.2 Nuclear Magnetic Resonance (NMR) spectroscopy

$^1$H NMR and $^{13}$C NMR spectroscopy were performed to confirm the purity of 2E1Br and 3,7D1OBr on an Agilent DDR2 500MHz NMR spectrometer using deuterated chloroform (CDCl$_3$) at room temperature. Chemical shifts were reported in parts per million (ppm) and were referenced using the residual $^1$H peak from the same deuterated solvent.

### 1.3 Differential Scanning Calorimeter (DSC)

The glass transition temperature, $T_g$, of all the samples were identified using a differential scanning calorimeter DSC 2500 (TA instruments). The instrument is equipped with a liquid nitrogen chiller pump to control the temperature. The DSC measurement were performed in the temperature range between 110K to 220K at a cooling rate of 10 K/min under a helium protective gas. A heating – cooling - heating cycle was employed to identify the glass transition. The $T_g$ values were determined from the midpoint of the specific heat capacity jump ($C_p$) upon the cooling cycle.

### 1.4 Density measurement

The mass density, $\rho$, of 2E1Br was measured through measurements of mass, $m$, and the corresponding volume, $V$, with $\rho = m/V$. Specifically, a fixed mass, $m$, of the 2E1Br was loaded onto a Anton Paar (MCR 302) rheometer with a pair of parallel plates of diameter of $D = 8\ mm$ and gap of $d = 1$ mm. Upon varying temperature from room temperature to its glass transition temperature, $T_g$, the mass of 2E1Br remains the same and the volume of the loaded liquid changes with temperature. The temperature was regulated by a CTD600 oven equipped with a liquid nitrogen evaporation unit (EVU20), which has an accuracy of ±0.1 K. To maintain the full loading between the parallel plate upon cooling, we manually reduce the gap $d$ between the top and bottom plate to compensate the thermal shrinkage of the liquid. The $d(T)$ at different temperatures is recorded through a digital camera that is independently calibrated by the gap at room temperature and by the size of the parallel plates at different temperatures. The volume of 2E1Br at low temperatures is $V(T) = \frac{\pi D^2 d(T)}{4}$ (See **Figure S1**). The mass density at different temperature, $\rho(T) = 4m/(\pi D^2 d(T))$.

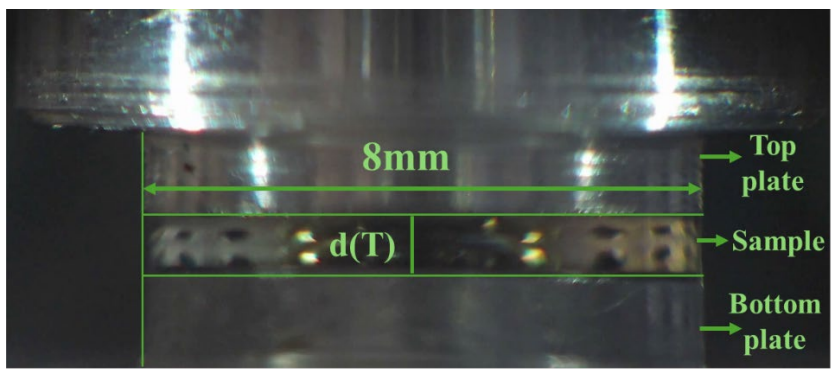

**Figure S1.** Image of the 2E1Br liquid at low temperatures in the oven between a pair of parallel plates with 8 mm in diameter. The gap at each temperature, $d(T)$, was estimated by knowing the digital diameter of the parallel plate.

## 2 Results

### 2.1 Nuclear Magnetic Resonance (NMR) spectroscopy

S3

Figures S2a and S2b show the ¹H NMR and ¹³C NMR of 2E1Br after distillation as well as the peak assignment. Figures S3a and S3b show the ¹H NMR and ¹³C NMR of 3,7D1OBr after the distillation and the peak assignment. The NMR results demonstrate the successful removal of impurities.

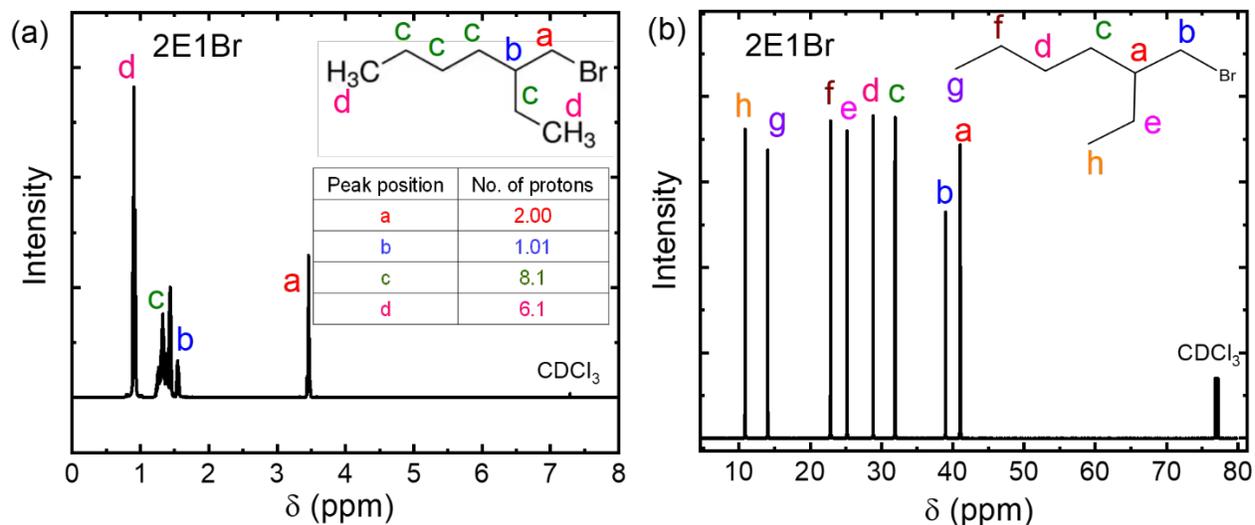

**Figure S2.** (a) ¹H-NMR and (b) ¹³C NMR of 2E1Br after purification.

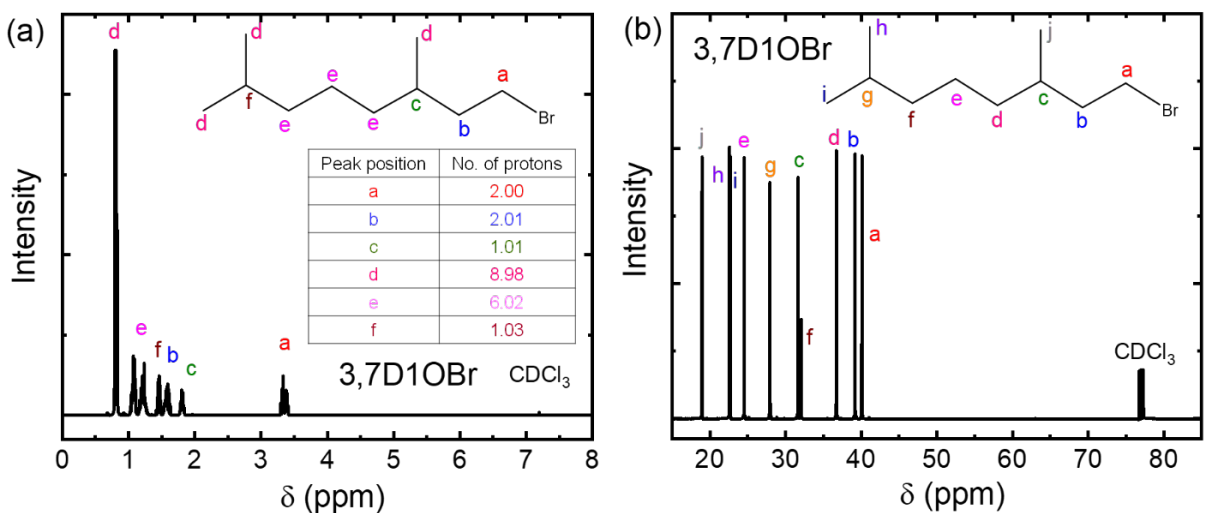

**Figure S3.** (a) ¹H-NMR and (b) ¹³C NMR of 3,7D1OBr after purification.

## 2.2 Broadband dielectric spectroscopy measurements of 2E1Br

**Figure S4** presents the derivative spectra, $\varepsilon'_{der}(\omega) = -\frac{\pi}{2}\frac{\partial \varepsilon'(\omega)}{\partial ln\omega}$, of 2E1Br before (open symbols) and after (filled symbols) purification, where $\varepsilon'(\omega)$ is the real part of the complex permittivity. The low-frequency shoulder peak of 2E1Br that is observed before purification does not appear after purification, indicating that the low-frequency shoulder peak originates from the impurities. According to the vendor, there are two types of impurities (~3-4%) in the 2E1Br: cyclohexane and potassium carbonate. To eliminate the influence of the impurities, all results in this work except the open symbols of **Figure S4** are obtained with purified 2E1Br and 3,7D1OBr.

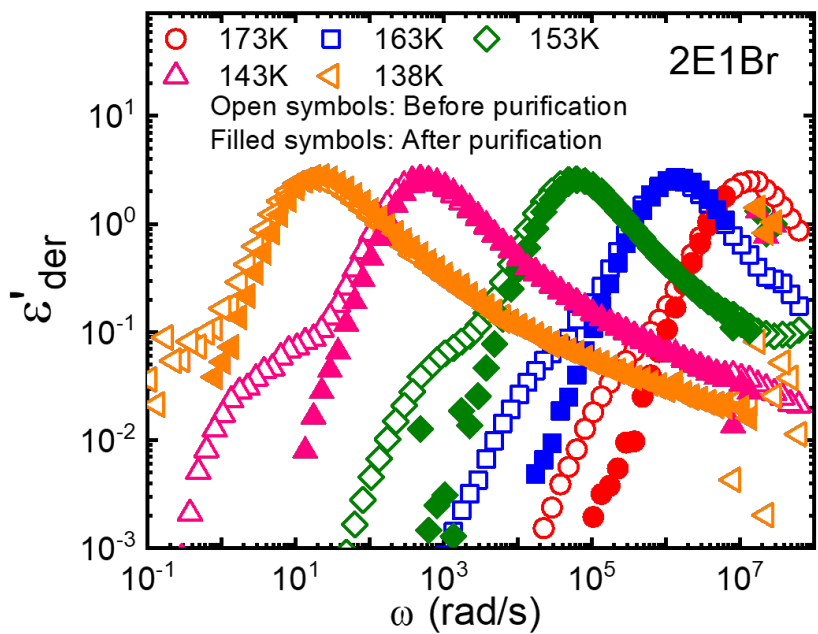

**Figure S4.** Derivative spectra, $\varepsilon'_{der}(\omega)$ at T = 173 K (red circles), 163 K (blue squares), 153 K (olive diamonds), 143 K (pink triangles, and 133 K (orange inverted triangles) of 2E1Br before (open symbols) and after purification (filled symbols).

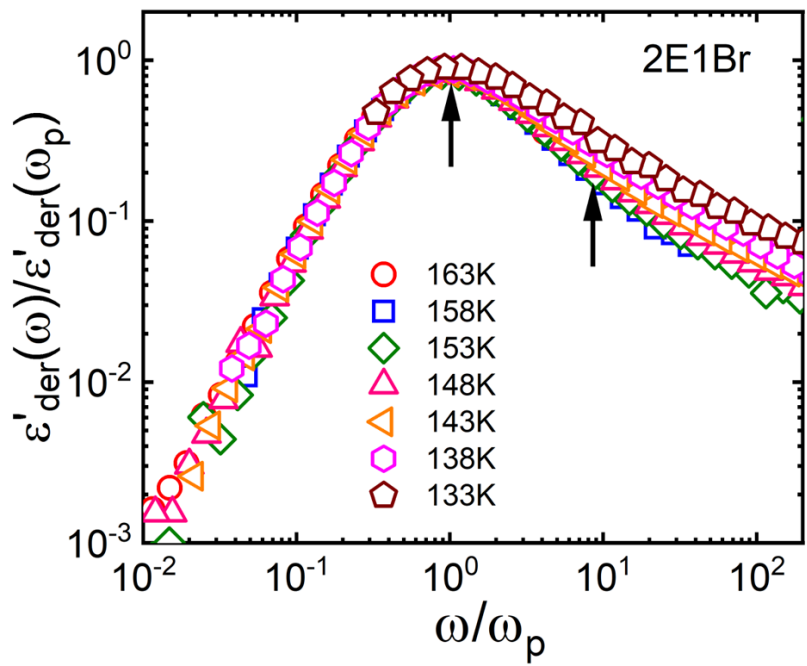

**Figure S5.** Normalized derivative spectra of purified 2E1Br. The arrows indicate the locations of the processes from the normalized derivative spectra.

## 2.3 Differential Scanning Calorimeter of 2E1Br

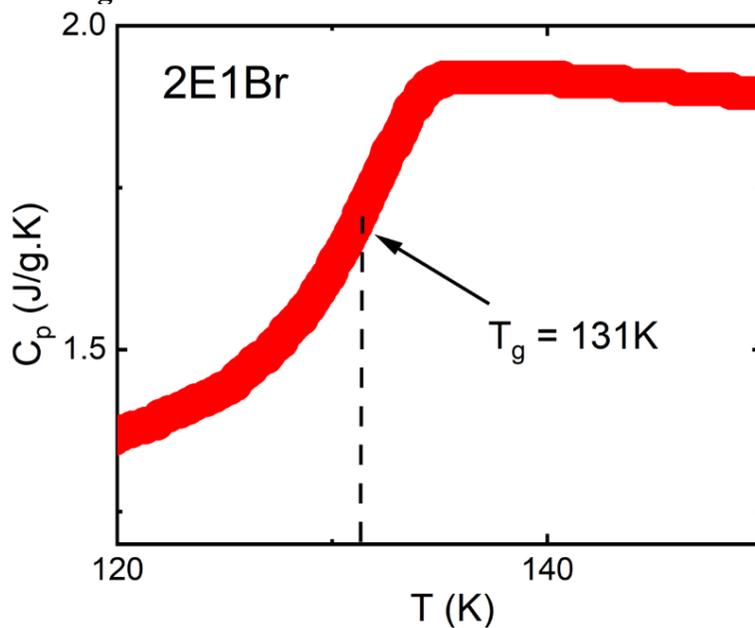

**Figure S6.** Specific heat capacity of 2E1Br. The dashed line represents the glass transition temperature ($T_g$) obtained from the midpoint of the glass transition step.

## 2.4 Rheology of 2E1Br

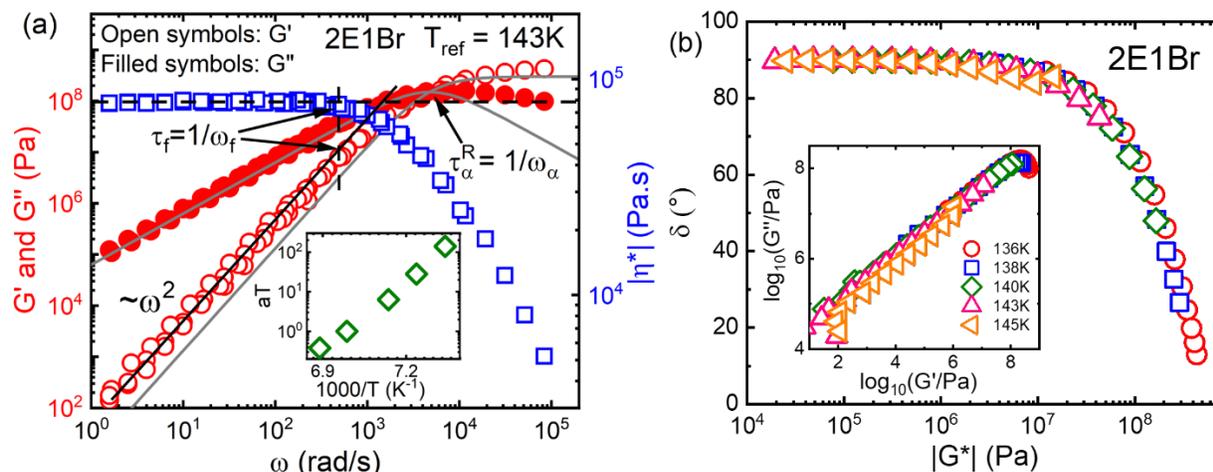

**Figure S7. (a)** Linear viscoelastic master curves of 2E1Br, where the open red circles represent the storage modulus, $G'$, and the filled circles represent the loss modulus, $G''$. The crossover between $G'$ and $G''$ highlight the location of the structural relaxation time $\tau_\alpha^R$. $G' \sim \omega^2$ holds at $\omega \leq \omega_f$, signifying the access of the terminal mode. The terminal relaxation time $\tau_f$ is defined through $\tau_f = 1/\omega_f$. At $\omega > \omega_f$, one also finds the amplitude of the complex viscosity, $|\eta^*|$, starts to deviate from its zero-shear value. The inset of panel a provides the dynamics shift factors $a_T$ of 2E1Br with respect to $T = 143K$. The solid grey lines are the corresponding calculations from the single-mode Maxwell model, $G'(\omega) = \frac{G_0(\omega\tau)^2}{1+(\omega\tau)^2}$ and $G''(\omega) = \frac{G_0\omega\tau}{1+(\omega\tau)^2}$, with characteristic relaxation time of $\tau = \tau_\alpha^R$ and a plateau modulus $G_0 = 2.3\times10^8\ Pa$. **(b)** The Van Gurp-Palmen plot to demonstrate the rheological simplicity of 2E1Br and the inset shows $log_{10}G''$ vs $log_{10}G'$.

## 2.5 SAXS measurement of 2E1Br

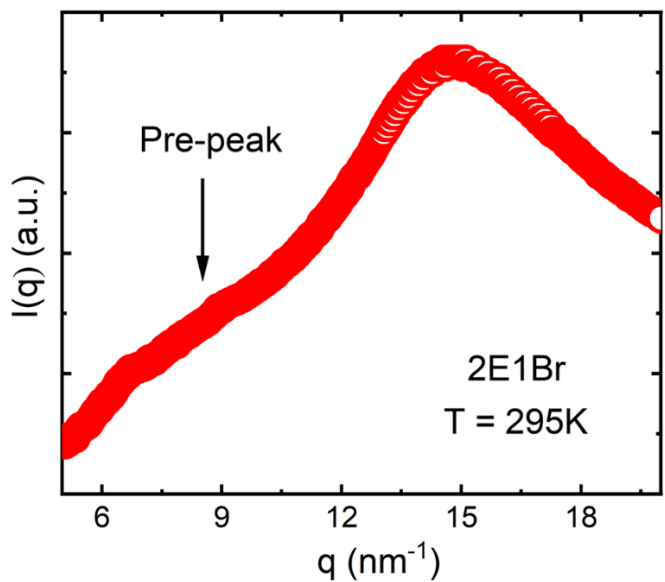

**Figure S8.** X-ray scattering intensity $I(q)$ of 2E1Br at 295K. Data is replotted from Ref.1

## 2.6 BDS of 3,7D1OBr and 2E1Cl

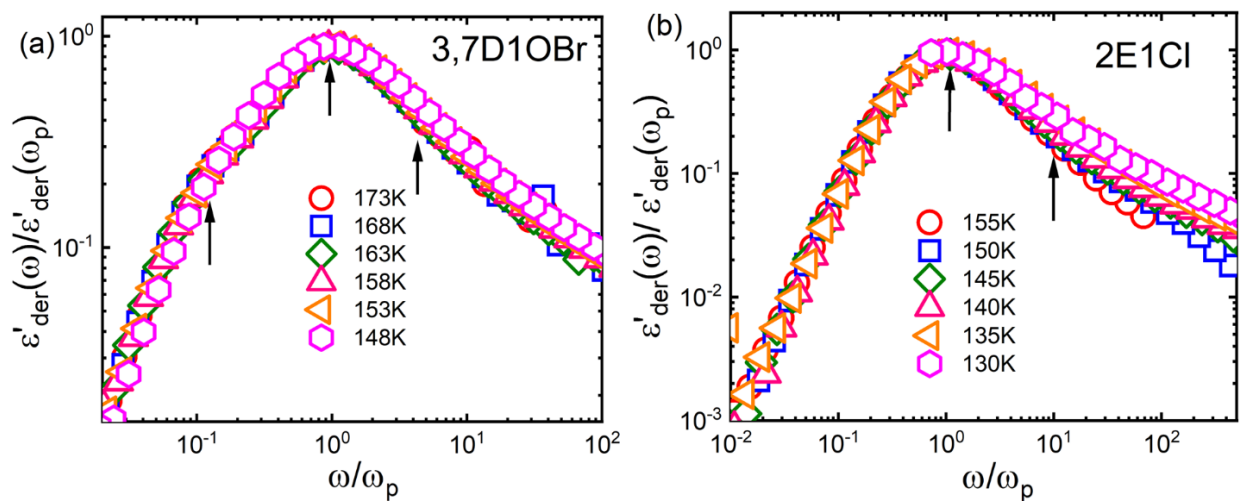

**Figure S9.** Normalized derivative spectra of **(a)** purified 3,7D1OBr and **(b)** 2E1Cl. Data is a reanalysis of dielectric data in Ref.1. The arrows indicate the locations of the processes from the normalized derivative spectra.

## 2.7 Rheology of 3,7D1OBr and 2E1Cl.

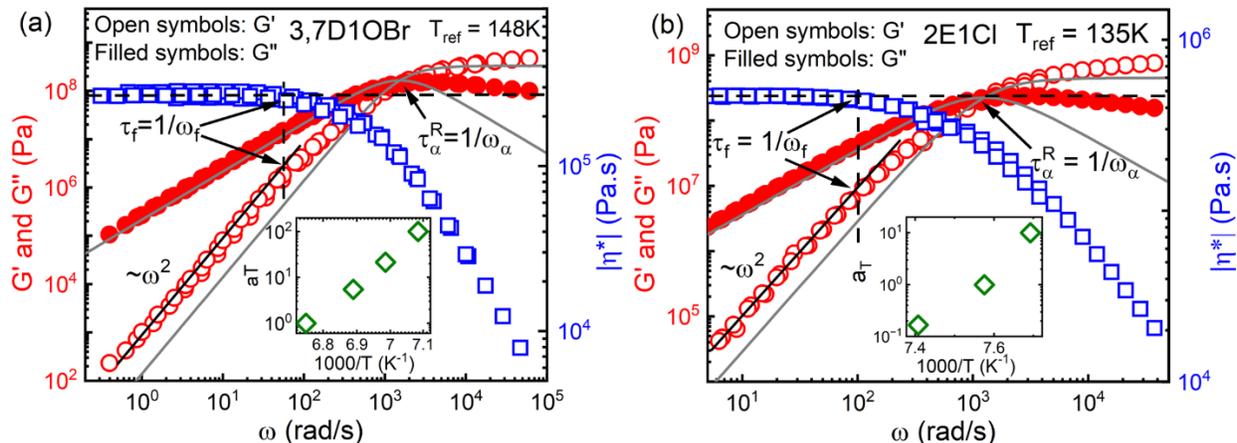

**Figure S10.** Linear viscoelastic master curves of **(a)** 3,7D1OBr and **(b)** 2E1Cl, where the open red circles represent the storage modulus, $G'$, and the filled circles represent the loss modulus, $G''$. The estimation of $\tau_\alpha^R$ and $\tau_f$ follow the same protocol as the one described in **Figure S6**. The open blue squares represent the amplitude of complex viscosity, $|\eta^*|$. The inset of panel **(a)** provides the shift factors, $a_T$ of 3,7D1OBr at reference temperature $T = 148 K$ and the inset of panel **(b)** gives the shift factors, $a_T$ of 2E1Cl at a reference temperature $T = 135\ K$. The solid grey lines of **Figure S7a** and **Figure S7b** are the single-mode Maxwell model prediction at $\tau = \tau_\alpha^R$ and $G_0 = 3.0 \times 10^8\ Pa$ for 3,7D1OBr and $\tau = \tau_\alpha^R$ and $G_0 = 5.0 \times 10^8\ Pa$ for 2E1Cl.

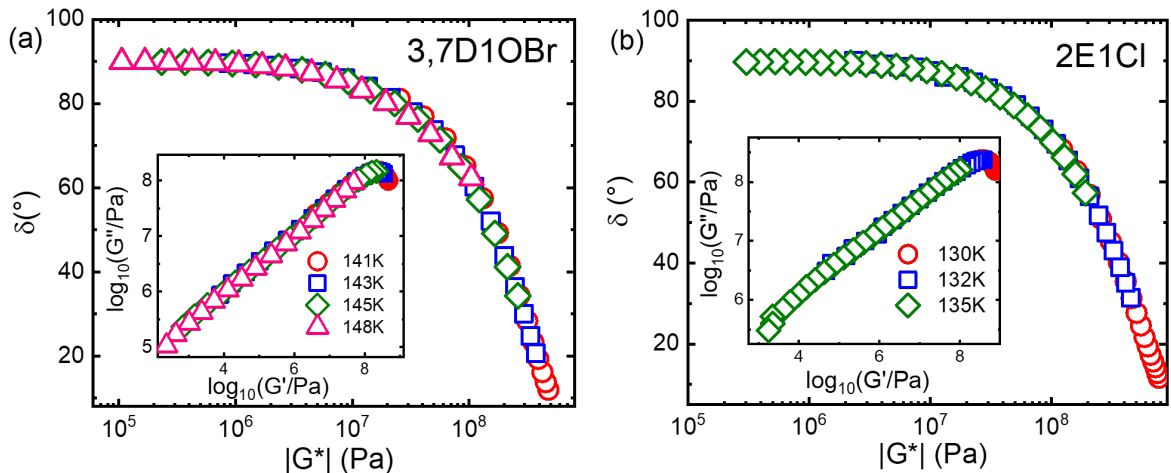

**Figure S11.** Van Gurp-Palmen plot showing the rheological simplicity of **(a)** 3,7D1OBr and **(b)** 2E1Cl. The inset of panels **(a)** and **(b)** provides the corresponding $\log_{10} G''$ vs $\log_{10} G'$.

## 2.8 Vogel-Fulcher-Tammann (VFT) fit parameters

Table S1. Vogel-Fulcher-Tammann (VFT) fit parameters

| Sample | Relaxation time | A | B | $T_0$(K) | $T_{100}$(K) |
|---|---|---|---|---|---|
| 2E1Br | $\tau_I$ | 12.8 ± 0.2 | 417.9 ± 22.2 | 102.3 ± 1.1 | 130 |
|  | $\tau_{II}$ | 14.4 ± 0.6 | 472.1 ± 53.8 | 99.1 ± 2.41 | 128 |
| 3,7D1OBr | $\tau_I$ | 12.1 ± 0.5 | 471.2 ± 59.5 | 103.4 ± 3.3 | 136 |
|  | $\tau_{II^*}$ | 12.3 ± 0.2 | 408.9 ± 23.9 | 107.3 ± 1.2 | 135 |
|  | $\tau_{II}$ | 11.5 ± 0.2 | 265.1 ± 14.7 | 115.1 ± 0.8 | 134 |
| 2E1Cl | $\tau_I$ | 13.8 ± 0.2 | 495.5 ± 25.7 | 93.8 ± 1.3 | 126 |
|  | $\tau_{II}$ | 12.1 ± 0.2 | 286.7 ± 15.6 | 103.5 ± 0.9 | 124 |

## 2.9 $g_k$ of 2E1Br.

Table S2. Values of $\varepsilon_\infty$, $\varepsilon_s$, and $\rho$ for $g_k$

| T (K) | $\varepsilon_\infty$ | $\varepsilon_s$ | $\rho \left(\frac{g}{cm^3}\right)$ | $g_k$ |
|---|---|---|---|---|
| 173 | 1.25 | 4.10 | 1.446 | 1.098 |
| 168 | 1.24 | 4.24 | 1.474 | 1.096 |
| 163 | 1.25 | 4.37 | 1.474 | 1.102 |
| 158 | 1.25 | 4.55 | 1.488 | 1.114 |
| 153 | 1.25 | 4.67 | 1.488 | 1.114 |
| 148 | 1.23 | 4.83 | 1.488 | 1.130 |
| 143 | 1.23 | 5.00 | 1.488 | 1.140 |
| 138 | 1.22 | 5.15 | 1.488 | 1.143 |

**Note:** The values of $\varepsilon_s$ are obtained from the dielectric storage permittivity $\varepsilon'(\omega)$ at low frequencies and $\varepsilon_\infty = \varepsilon_s - \Delta\varepsilon_I - \Delta\varepsilon_{II}$ from the HN function fit. Additionally, the values of $M = 193.12$ and $\mu = 1.48$ were used.

## 3 References

1. Büning, T.; Lueg, J.; Bolle, J.; Sternemann, C.; Gainaru, C.; Tolan, M.; Böhmer, R., Connecting structurally and dynamically detected signatures of supramolecular Debye liquids. *J. Chem. Phys.* **2017**, *147* (23), 234501.